\documentclass[a4paper]{article}              % Book class in 11 points

\pdfoutput=1
\usepackage[pdftex]{graphicx,color}
\usepackage[pdftex]{color}
\usepackage[colorlinks,bookmarks]{hyperref}
\usepackage{multirow}
\usepackage[export]{adjustbox}
\definecolor{linkblue}{rgb}{0,0,0.8}
\definecolor{linkgreen}{rgb}{0,0.5,0}
\hypersetup{linkcolor=linkblue, citecolor=linkgreen, urlcolor=linkblue}

\usepackage[T1]{fontenc}
\usepackage[utf8]{inputenc}
\usepackage{lmodern}
\usepackage[english]{babel}
\usepackage{url}
\usepackage{bm}
\usepackage[us]{datetime}
\usepackage{booktabs}
\usepackage{enumerate}
\usepackage{graphics}
\usepackage{float}
\usepackage{amsmath}
\usepackage{amsfonts}
\usepackage{amssymb}

\usepackage{authblk}

\usepackage{natbib}
\usepackage{etoolbox}

\graphicspath{{./figs/}}

\newcommand{\pa}[1]{\left(#1\right)}
\newcommand{\be}{\begin{equation}}
\newcommand{\ee}{\end{equation}}
\newcommand{\ba}{\begin{array}}
\newcommand{\ea}{\end{array}}
\newcommand{\ds}{\displaystyle}

\newcommand{\jpas}{\mbox{J-PAS}\ }

\title{Observing the dark sector\\
\ \\
\small{Contribution to the 3rd José Plínio Baptista School on Cosmology
``The Dark Sector of the Universe''
held in 2016 in Pedra Azul, Espírito Santo, Brazil}
}

\author[1]{Valerio Marra}
\author[2,3]{Rogerio Rosenfeld}
\author[4]{Riccardo Sturani}

\affil[1]{\small Núcleo Cosmo-ufes \& Departamento de Física,
Universidade Federal do Espírito Santo, 29075-910, Vitória-ES, Brazil}

\affil[2]{\small ICTP South American Institute for Fundamental Research \& 
Instituto de Física Teórica, Universidade Estadual Paulista, 01140-070, São Paulo-SP, Brazil}

\affil[3]{\small Laboratório Interinstitucional de e-Astronomia - LIneA, 20921-400, Rio de Janeiro-RJ, Brazil}

\affil[4]{\small International Institute of Physics, Universidade Federal do Rio Grande do Norte,
Campus Universitario, Lagoa Nova, Natal-RN 59078-970, Brazil}

\date{}

\begin{document}

\maketitle

\begin{abstract}
Despite the observational success of the standard model of cosmology, present-day observations do not tightly constrain the nature of dark matter and dark energy and modifications to the theory of general relativity. Here, we will discuss some of the ongoing and upcoming surveys that will revolutionize our understanding of the dark sector.
\end{abstract}

\tableofcontents

\section{Introduction}

The standard model of cosmology is of utter simplicity: assuming General Relativity and small perturbations about a spatially homogeneous and isotropic background model, it can easily account for basically all cosmological observations, probing a vast range of scales in space and time with just six parameters. According to the standard model of cosmology the universe is dominated by a mysterious matter called ``dark matter'' and a mysterious energy called ``dark energy''.
This conclusion is supported, for example, by observations of supernovae Ia (SNIa)~\citep{Betoule:2014frx,Scolnic:2017caz}, of the Baryonic Acoustic Oscillations (BAO)~\citep{Kazin:2014qga,Alam:2016hwk}, of the anisotropies of the Cosmic Microwave Background (CMB)~\citep{Hinshaw:2012aka,Aghanim:2018eyx} and of the weak lensing of galaxies \citep{Kohlinger:2017sxk,Abbott:2017wau}.
Dark matter seeded the formation of galaxies while dark energy  is driving them apart by causing the universe to accelerate, a phenomenon that was conjectured in the early 1990s \citep{Krauss:1995yb}, observed in 1998 \citep{Perlmutter:1998np,Riess:1998cb} and awarded the Nobel Prize in 2011.

However, a satisfactory theoretical explanation of dark matter and dark energy -- the so-called dark sector -- is still lacking  and their properties  are not yet well constrained by the data~\citep{Feng:2010gw,Li:2011sd,Capozziello:2019cav}.
According to the standard model of cosmology, baryons -- particles belonging to the successful standard model of particle physics -- constitute only 5\% of the energy content of the universe.
The remaining 95\% is left to the dark sector. Approximately 25\% consists of a yet-undetected matter component, which is thought to be a massive particle of non-baryonic nature which interacts through gravity and weak interaction only.
It is named ``cold dark matter'' because
it neither emits nor absorbs light or other electromagnetic radiation 
(and so it is dark) and it moves slowly compared to the speed of light (and so it has a low temperature).
Dark energy is responsible for the missing 70\%. The best candidate to date is the ``cosmological constant'', the energy of the vacuum and  an arbitrary constant of nature in the general relativity. Its fundamental property -- gravitational repulsion for positive energy density --
causes the expansion of the universe to accelerate, as mentioned earlier.

Therefore, the standard model of cosmology is facing a formidable challenge as it is asked to account for not one but two  unknown components.
It is undeniable that cosmology itself is at the moment built on shaky foundations, relying on an unexplained dark sector for observations to fit the model. The implications of this cannot be overstated.
This is the motivation for the large theoretical and experimental effort that is being deployed to better understand the nature of the dark sector.

Indeed, the scientific study of the universe is on the verge of a revolution. New cosmological galaxy surveys will map the universe in unprecedented detail over volumes which we have only been able to imagine in vast computer simulations.
Gravitational-wave survey will soon produce massive catalogs of events able to probe the theory of General Relativity in the uncharted strong-field limit
and determine the presently widely unknown stellar black hole mass function, besides providing a new handle to measure late cosmological acceleration.
Finally,  21-cm survey will map the density of the universe at even larger scales, covering almost all the observable universe.

In the following we will summarize how these surveys will revolutionize our understanding of the dark sector, pinning down the phenomenology of dark matter and dark energy and so triggering major progress in our understanding of the fundamental interactions of nature.

%%%%%%%%%%%%%%%%%%%%%%%%%%%%%%%%%%%%%%%%%%%%%%%%%%%%%%%%%%%%%%%%
\section{Galaxy surveys} \label{galsur}

\begin{figure}
\centering
\includegraphics[width=\textwidth]{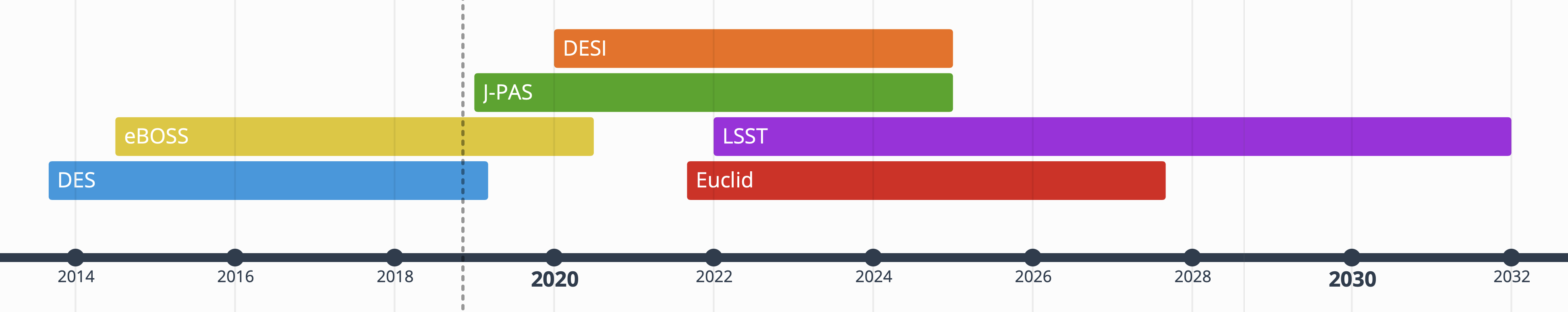}
\caption{Timeline of the ongoing and upcoming galaxy surveys discussed in Section~\ref{galsur}.\label{fig:gs-timeline}}
\end{figure}

%\textcolor{red}{\bf [VM: verify!]}
Figure \ref{fig:gs-timeline} shows the timeline of the ongoing and upcoming galaxy surveys that will be discussed in this section.
Surveys have been classified according to their constraining power on the dark energy equation of state \citep{Albrecht:2006um}:
\begin{equation} \label{weq}
w = \frac{p}{\rho} \,,
\end{equation}
where $p$ is the pressure of the fluid and $\rho$ its energy density. Dark energy has been constrained to have an equation of state of $w\approx -1$, while dark matter has been found compatible with $w\approx 0$, a pressureless (dust) fluid.%
\footnote{See \citet{Luongo:2018lgy} for the case of dark matter with a non-vanishing pressure.}
If $w=-1$, one has the cosmological constant and the $\Lambda$CDM model. If one lets $w$ free, then one has the $w$CDM model.

The constraining power on the dark energy equation of state is quantified via the so-called Figure of Merit (FoM), which is defined as
\begin{equation}
\text{FoM}={\det}^{-1}F(w_0, w_a) \,,
\end{equation}
where $F$ is the (marginalized) Fisher matrix relative to the dark energy equation of state, parameterized according to%
\footnote{See \citet{Aviles:2012ay} for alternative parametrizations.}
\begin{equation}
w(a)=w_0+(1-a)w_a \,.
\end{equation}
Stage-II experiments (previous to DES) feature a FoM less than 50, Stage-III experiments about 50--200 and Stage-IV experiments about 200+. %Planck and Stage-II priors are adopted.
DES and eBOSS are Stage-III level, \jpas approches Stage IV and the remaining are all ``Stage IV''.

%https://time.graphics/line/168966
%%%%%%%%%%%%%%%%%%%%%%%%%%%%%%%%%%%%%%%%%%%%%%%%%%%
%%%%%%%%%%%%%%%%%%%%%%%%%%%%%%%%%%%%%%%%%%%%%%%%%%%%%

\subsection{Extended Baryon Oscillation Spectroscopic Survey}

\begin{figure}
\centering
\includegraphics[width=9
cm]{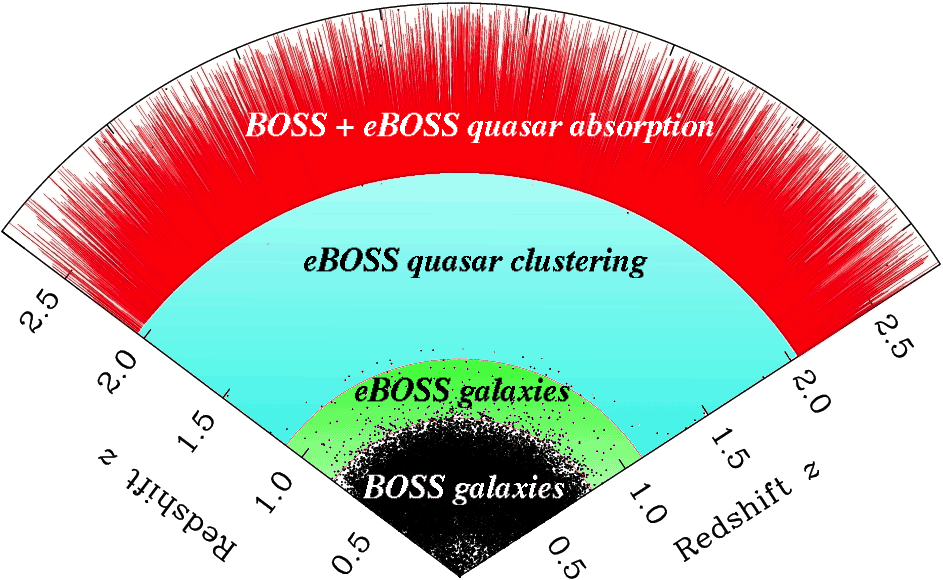}
\caption{eBOSS survey as compared with the BOSS survey. From \href{https://www.sdss.org/surveys/eboss/}{sdss.org/surveys/eboss}.}
\label{fig:eBOSS}
\end{figure}

The Extended Baryon Oscillation Spectroscopic Survey (eBOSS) is part of the fourth phase of the Sloan Digital Sky Survey (SDSS-IV) and extends the Baryon Oscillation Spectroscopic Survey (BOSS, part of SDSS-III) to much higher redshifts.
The eBOSS survey started on July 2014 and will last 6 years, and will produce the largest volume survey to date, see Figure~\ref{fig:eBOSS}.
eBOSS targets the observation of galaxies and  quasars in a range of redshifts currently left uncharted by other maps of the large-scale structure of the universe.

%%%%%%%%%%%%%%%%%%%%%%%%%%%%%%%%%%%%%%%%%%%%%%%%%%%
%%%%%%%%%%%%%%%%%%%%%%%%%%%%%%%%%%%%%%%%%%%%%%%%%%%%%

300,000 luminous red galaxies (LRG) will be observed over 7500 deg$^2$ in the redshift range $0.6 < z < 0.8$,
189,000 emission line galaxies (ELG) over 1000 deg$^2$ in the range $0.6 < z < 1.0$
and 573,000 quasars over 7500 deg$^2$ in the range $0.9 < z < 3.5$. This large catalog will produce 1-2\% distance measurements from baryon acoustic oscillations in the redshift range $0.6 < z < 2.5$.
At this time, the Data Release 14 (DR14) of the first 2 years of observations has been publicly released, see \cite{Ata:2017dya} for the first measurement of baryon acoustic oscillations between redshift 0.8 and 2.2.
See \cite{Blanton:2017qot} for further information.

\subsection{Dark Energy Survey}
The Dark Energy Survey (DES)\footnote{www.darkenergysurvey.org} is a project that is mapping 5000 deg$^2$ of the sky (approximately 1/8 of the whole sky) using 525 nights of observations in 5 years at the Blanco Telescope in the Cerro Tololo Inter-American Observatory in Chile. The project is led by Fermilab, a US  national laboratory near Chicago, and its current (third) Director is Rich Kron from the University of Chicago. There are more than 400 scientists from over 25 institutions in the US, Brazil, Spain, UK, Germany, Switzerland and Australia working on the project.

A large digital camera with 570 Megapixels in 62 CCD’s was built by the collaboration and installed in the telescope. This so-called DECam takes exposures using 5 filters (grizY) that provide an estimate of the photometric redshift of approximately 300 million objects. This large amount of data is transferred and  processed at the National Center of Supercomputing Applications (NCSA) in Urbana-Champaign to generate a value-added catalogue.

The first light of DES was in 2012. There was a 6-month extension to the observation period that ends in January 2019. There are already more than 200 papers from the DES collaboration in the Inspire database. Results from the first year of observations have been published leading to several ground-breaking results, some of which will be mentioned below. Some highlights are:
\begin{itemize}
\item Produced the largest contiguous mass map of the Universe;
\item Discovered nearly a score of Milky Way dwarf satellites and other Milky Way structures; 
\item Measured weak lensing cosmic shear, galaxy clustering, and cross-correlations with CMB lensing, and with clusters detected via X-ray and the Sunyaev-Zeldovich effect; 
\item  Measured light curves for large numbers of type Ia supernovae and discovered a number of super-luminous supernovae (SLSN) including the highest-redshift SLSN so far; 
\item  Discovered a number of redshift $z>6$ quasars (also known as QSOs or quasi-stellar objects); 
\item  Discovered a number of strongly lensed galaxies and QSOs;
\item Discovered a number of interesting objects in the outer Solar System;
\item  Found optical counterparts of GW events% – led by a brazilian who studied in UFES - Marcelle Soares-Santos.
\end{itemize}

\begin{figure}
\centering
\includegraphics[width=8.5cm]{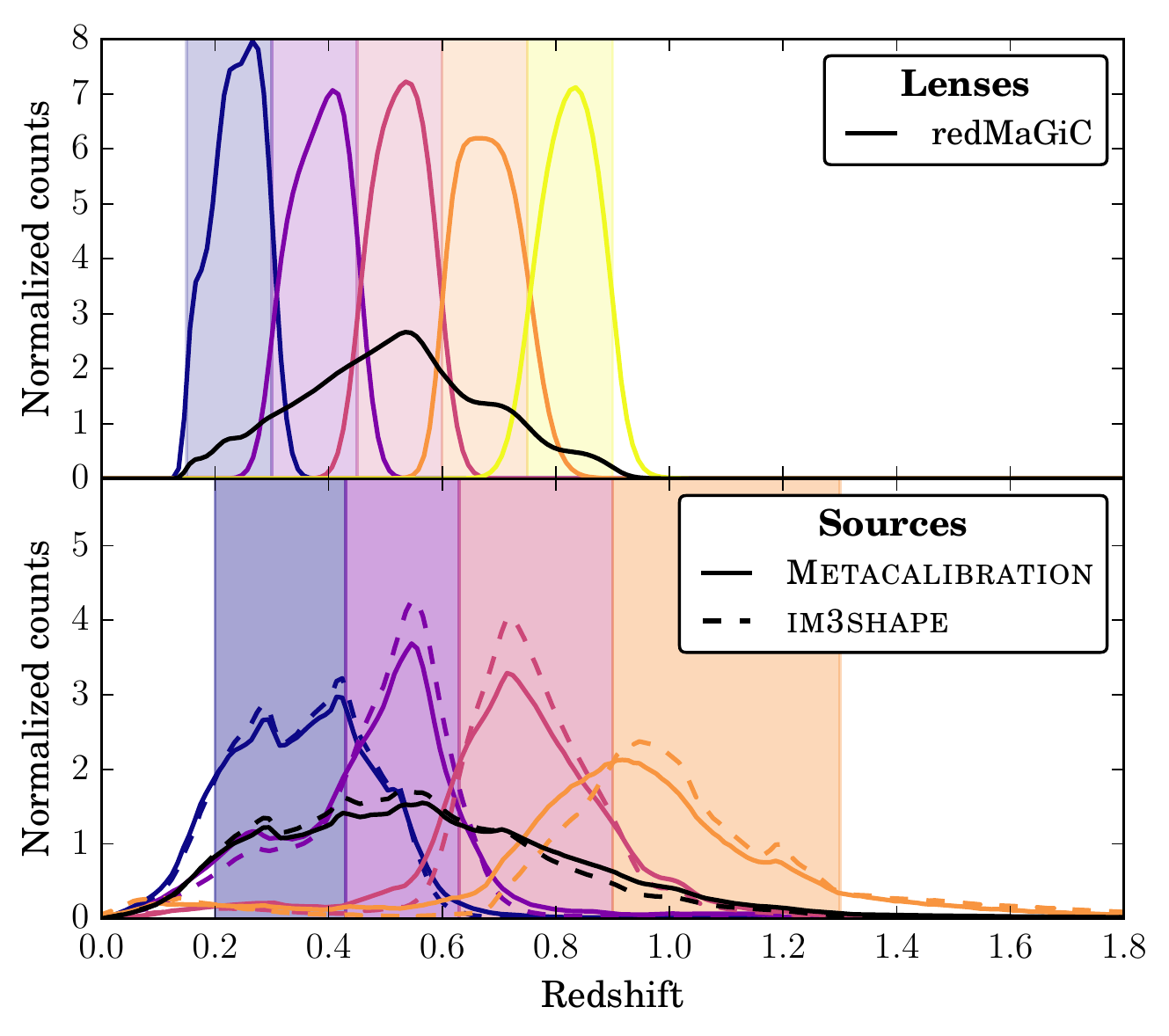}
\caption{Photometric redshift distributions for the galaxy position (lens) and  shear (sources) catalogs.
The shaded regions mark the redshift bins: galaxies are divided according to their mean photo-z estimate. The redshift distributions of galaxies in each bin is shown with colored lines, while their overall redshift distributions with black lines. 
While the lens galaxies are analyzed using only one pipeline (redMaGiC), source galaxies are analyzed with two pipelines (IM3SHAPE and METACALIBRATION). From \citet{Abbott:2017wau}.
}
\label{fig:RedShiftDistribution}
\end{figure}

DES combines four different observational probes in order to find the best constraints on dark energy:
\begin{itemize}
\item Distribution of ~300 million galaxies, including measurements of the Baryon Acoustic Oscillation;
\item Weak gravitational lensing of galaxies; 
\item Supernovae of type Ia;
\item Counts of clusters of galaxies.
\end{itemize}
The main cosmological result of the first year of observations was published in \cite{Abbott:2017wau}, a key paper which uses results of other 11 papers. It combines measurements of three 2-point correlation functions involving galaxy positions and weak lensing (shear): galaxy-galaxy (galaxy clustering), galaxy-shear and shear-shear.
Two galaxy samples are used:
\begin{itemize}
\item ``Shape catalogue'': 26M galaxies for cosmic shear
measurements (source galaxies) divided into 4
redshift bins;
\item ``Position catalogue'': 650,000 luminous red galaxies
(lens galaxies) for clustering measurements divided
into 5 redshift bins.
\end{itemize}
The photometric redshift distributions for the two samples are shown in Fig.~\ref{fig:RedShiftDistribution}.

%%%%%%%%%%%%%%%%%%%%%%%%%%%%%%%%%%%%%%%%%%%%%%%%%%%
%%%%%%%%%%%%%%%%%%%%%%%%%%%%%%%%%%%%%%%%%%%%%%%%%%%%%

\begin{figure}
\centering
\includegraphics[width=\textwidth]{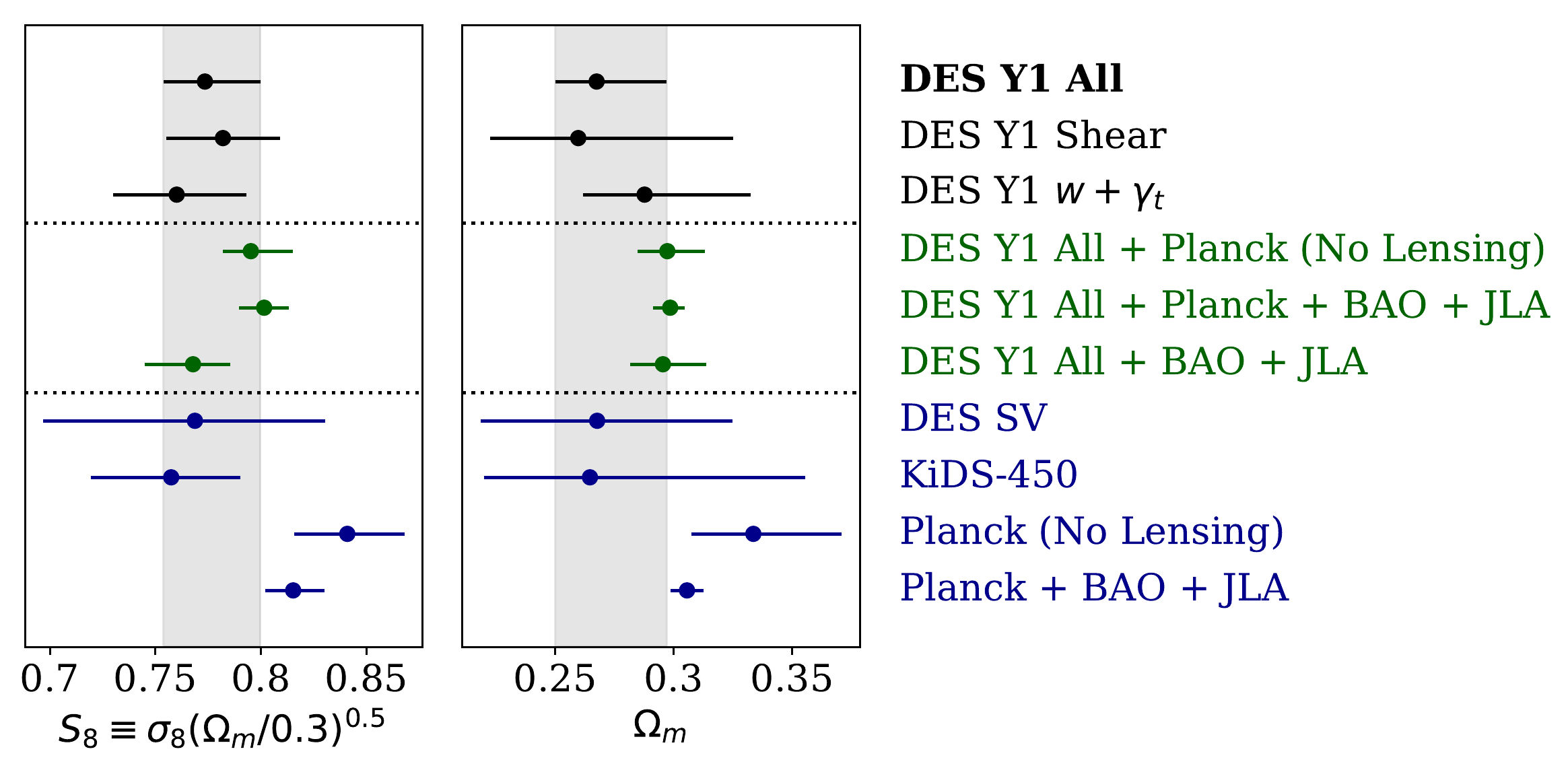}
\caption{Results for $S_8$ and $\Omega_m$ within $\Lambda$CDM.
DES Y1 refers to the first year of observations of DES, DES SV to the  Science Verification analysis, KiDS-450 to the weak lensing analysis from the Kilo Degree Survey \citep{Kohlinger:2017sxk}, Planck and JLA to CMB and supernova Ia analyses, respectively. From \citet{Abbott:2017wau}.}
\label{fig:PullLCDM}
\end{figure}

The data vectors were defined using scale cuts to mitigate non-linear bias effects and it comprises 457 entries (different redshift bins, angular bins, correlation functions). We used a theoretical (halo-model based) covariance matrix (dimension 457x457) computed with the CosmoLike code  validated with 800 lognormal mocks.
For the Markov Chain Monte Carlos (MCMC) analysis we had 20 nuisance parameters (related to the redshift uncertainty, galaxy bias, intrinsic alignment and shear calibration) in addition to the usual 6 cosmological parameters for the spatially flat $\Lambda$CDM model (7 for $w$CDM, where $w$ is defined in equation \eqref{weq}). We concentrate the analysis on the two most sensitive parameters: $\Omega_m$ and $S_8=\sigma_8 (\Omega_m/0.3)^{0.5}$, where $\Omega_m$ is the matter density parameter and $\sigma_8$ is the root mean square mass fluctuation on a scale of $8h^{-1}$ Mpc.
The matter density parameter is defined according to $\Omega_m= \rho_{m0} /\rho_{c0}$, where $\rho_{m0}$ is the present-day matter density and the present-day critical density is given by $\rho_{c0}=3 H_0^2/(8 \pi G)$, where $H_0$ is the Hubble-Lemaître constant and $G$ is Newton's gravitational constant.
We also compare results from DES alone with DES combined with data such as CMB, BAO and SNIa, see Fig.~\ref{fig:PullLCDM}.

%%%%%%%%%%%%%%%%%%%%%%%%%%%%%%%%%%%%%%%%%%%%%%%%%%%
%%%%%%%%%%%%%%%%%%%%%%%%%%%%%%%%%%%%%%%%%%%%%%%%%%%%%

\begin{figure}
\centering
\includegraphics[width=9 cm]{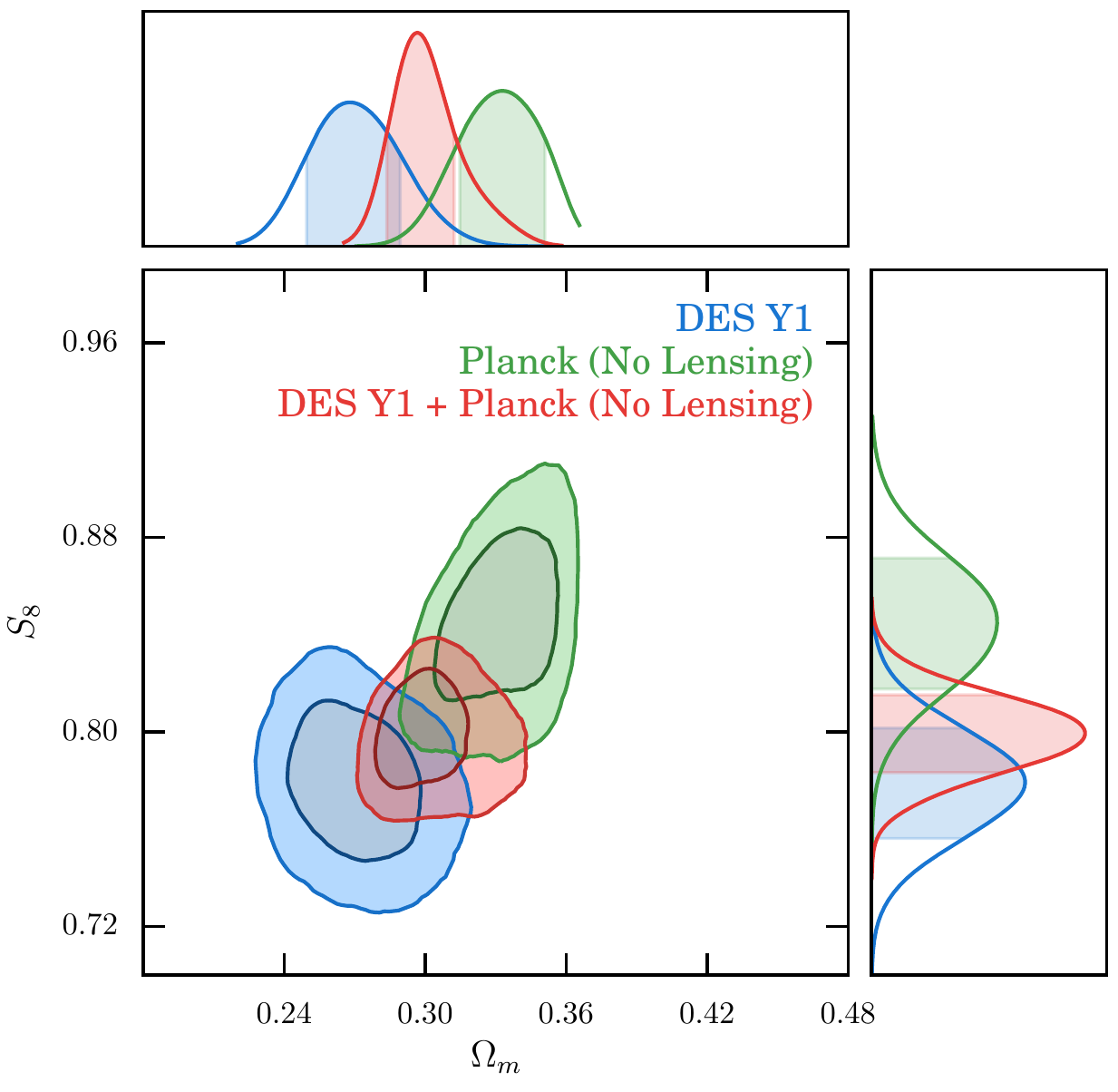}
\caption{The contours show the 1- and 2-$\sigma$ constraints for $S_8$ and $\Omega_m$ within $\Lambda$CDM. The shaded area in the 1-d posteriors shows the 68\% confidence region. From \citet{Abbott:2017wau}.}
\label{fig:S8Om}
\end{figure}

In Fig.~\ref{fig:S8Om} we show the 1- and 2-$\sigma$ contours for the parameters  $S_8$ and $\Omega_m$ obtained from DES, Planck and combined. It's amazing to see that, for the first time, results from large surveys of galaxies provide bounds on cosmological parameters that are competitive with the ones obtained from CMB. It also shows the consistency of the $\Lambda$CDM model from the time of recombination where the CMB was generated to late times after galaxy formation.

%%%%%%%%%%%%%%%%%%%%%%%%%%%%%%%%%%%%%%%%%%%%%%%%%%%
%%%%%%%%%%%%%%%%%%%%%%%%%%%%%%%%%%%%%%%%%%%%%%%%%%%%%

\begin{figure}
\centering
\includegraphics[width=\textwidth]{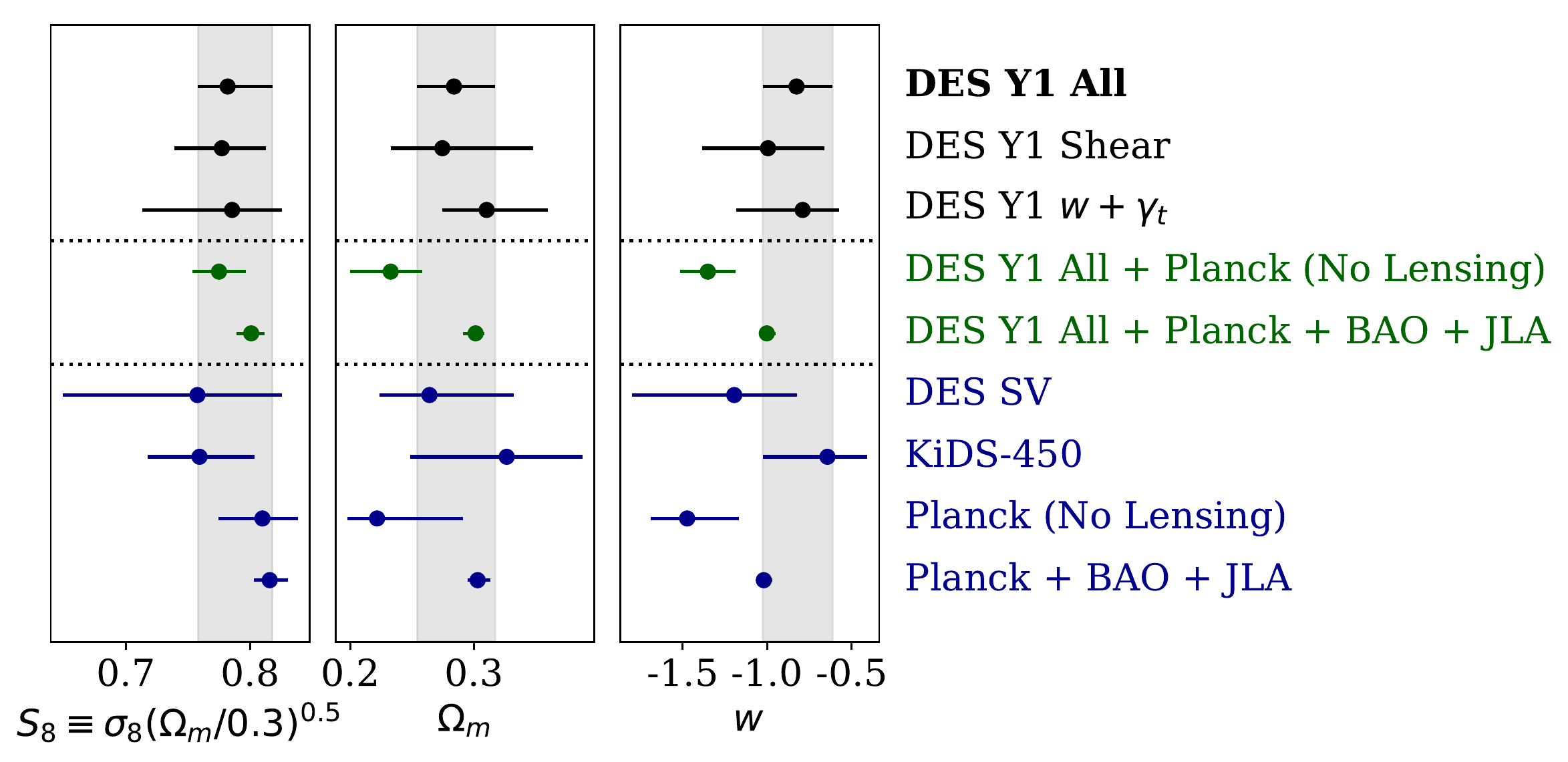}
\caption{Results for $S_8$ and $\Omega_m$ within $w$CDM obtained from DES and other experiments, similar to Figure~\ref{fig:PullLCDM}. From \citet{Abbott:2017wau}.}
\label{fig:PullwCDM}
\end{figure}

DES data were analyzed also in the context of the $w$CDM model which features a constant equation of state $w$, see Fig.~\ref{fig:PullwCDM}. The result for $w$, when DES is combined with other data, provides the state-of-the-art determination of $w$ \citep{Abbott:2017wau}:
\begin{equation}
    w = -1.00 ^{+0.05}_{-0.04} \,,
\end{equation}
in perfect agreement with $\Lambda$CDM.

%Pull for wCDM
%%%%%%%%%%%%%%%%%%%%%%%%%%%%%%%%%%%%%%%%%%%%%%%%%%%
%%%%%%%%%%%%%%%%%%%%%%%%%%%%%%%%%%%%%%%%%%%%%%%%%%%%%

\begin{figure}
\centering
\includegraphics[width=7 cm]{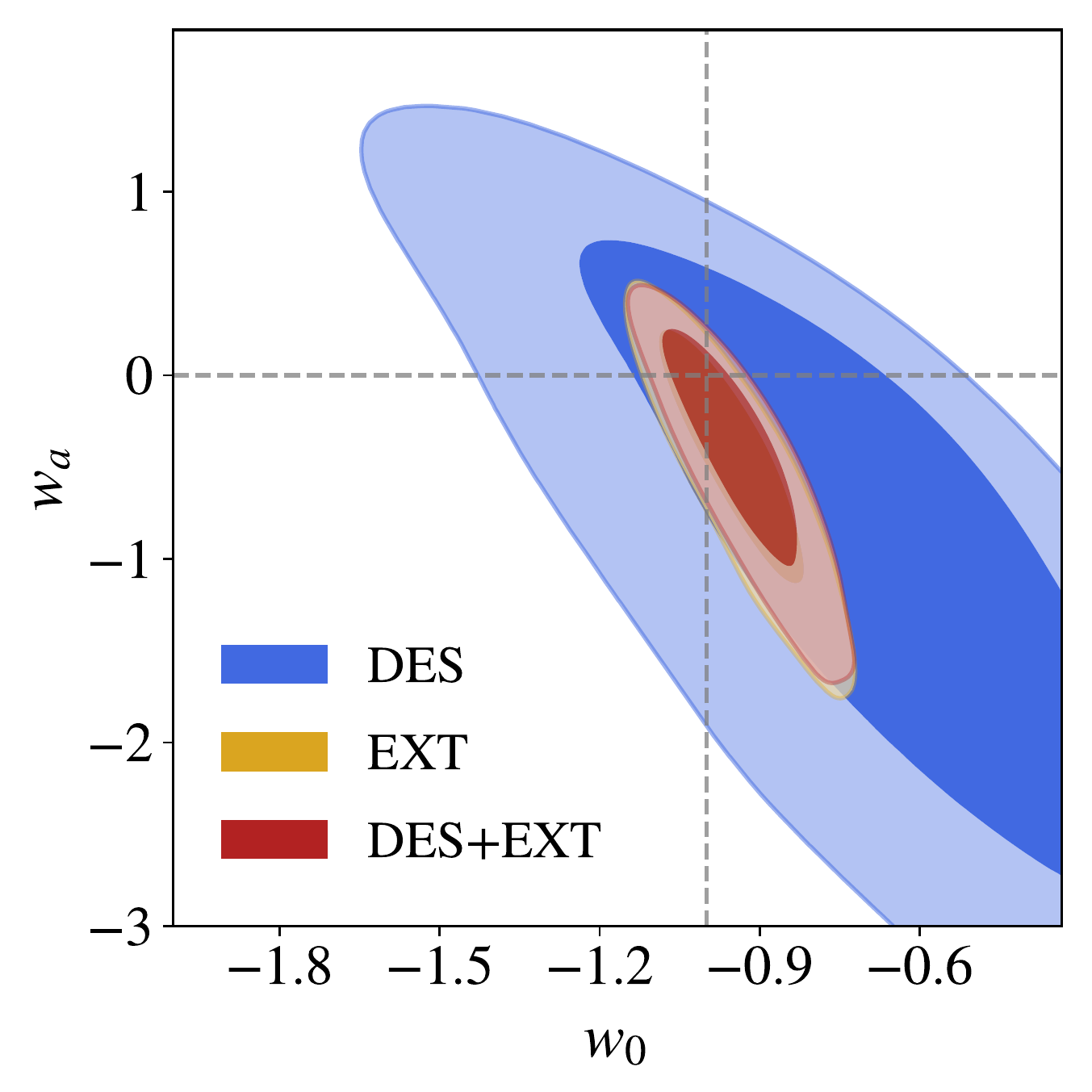}
\caption{
The contours show the 1- and 2-$\sigma$ constraints for the dark energy equation of state parameters $w_0$ and $w_a$ of equation~\eqref{weq}  obtained from DES and other experiments. From \citet{Abbott:2018xao}.}
\label{fig:w0wa}
\end{figure}

Other extensions of the $\Lambda$CDM model were studied in \citet{Abbott:2018xao}:
\begin{itemize}
\item Spatial curvature;
\item The effective number of neutrinos species;
\item Time-varying equation of state of dark energy, see equation \eqref{weq}; 
\item Tests of gravity.
\end{itemize}
As an example, in Fig.~\ref{fig:w0wa} we show the contour plots for $w_0$ and $w_a$ for DES and other external data. We can see that DES data from the first year of observation is still not competitive with other data.

%Contours for w0-wa
%%%%%%%%%%%%%%%%%%%%%%%%%%%%%%%%%%%%%%%%%%%%%%%%%%%
%%%%%%%%%%%%%%%%%%%%%%%%%%%%%%%%%%%%%%%%%%%%%%%%%%%%%

\begin{figure}
\centering
\includegraphics[width=9.5 cm]{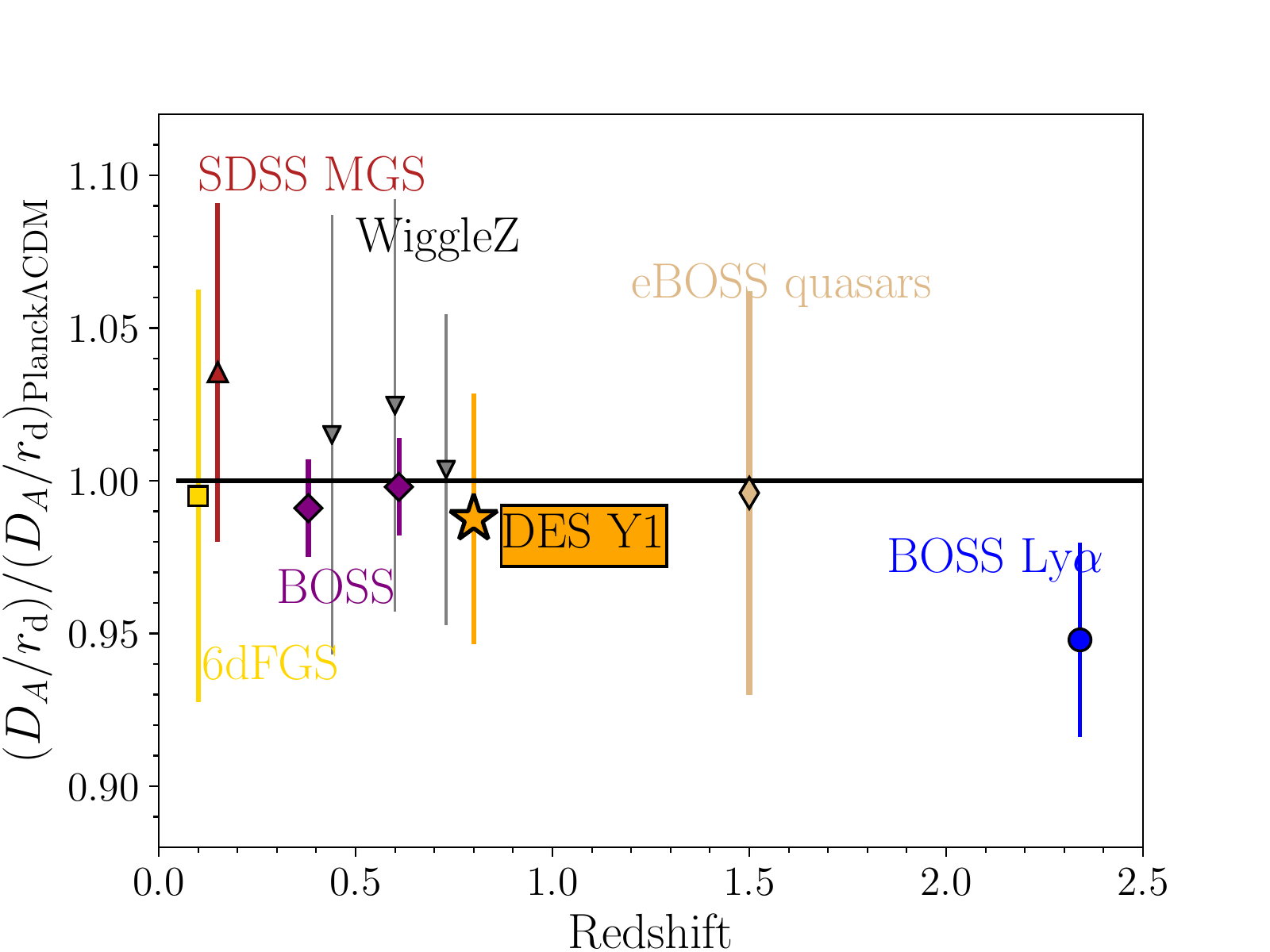}
\caption{
Measurement of the angular diameter distance from DES, compared to the Planck prediction and other measurements. From \citep{Abbott:2017wcz}.}
\label{fig:BaoDES}
\end{figure}

The DES data also produced the measurement of what is called the shift parameter $\alpha$ which gives the location of the BAO peak with respect to a reference cosmology \citep{Abbott:2017wcz}.
In Fig.~\ref{fig:BaoDES} we show the DES measurement of the angular diameter distance $D_A$, corresponding to the BAO feature, compared to other measurements at different redshifts.

%BAO DES
%%%%%%%%%%%%%%%%%%%%%%%%%%%%%%%%%%%%%%%%%%%%%%%%%%%
%%%%%%%%%%%%%%%%%%%%%%%%%%%%%%%%%%%%%%%%%%%%%%%%%%%%%

\subsection{Javalambre Physics of the Accelerating Universe Astrophysical Survey}

%%%%%%%%%%%%%%%%%%%%%%%%%%%%%%%%%%%%%%%%%%%%%%%%%%%
\begin{figure}
\centering %trim={<left> <lower> <right> <upper>}
\includegraphics[width=\textwidth]{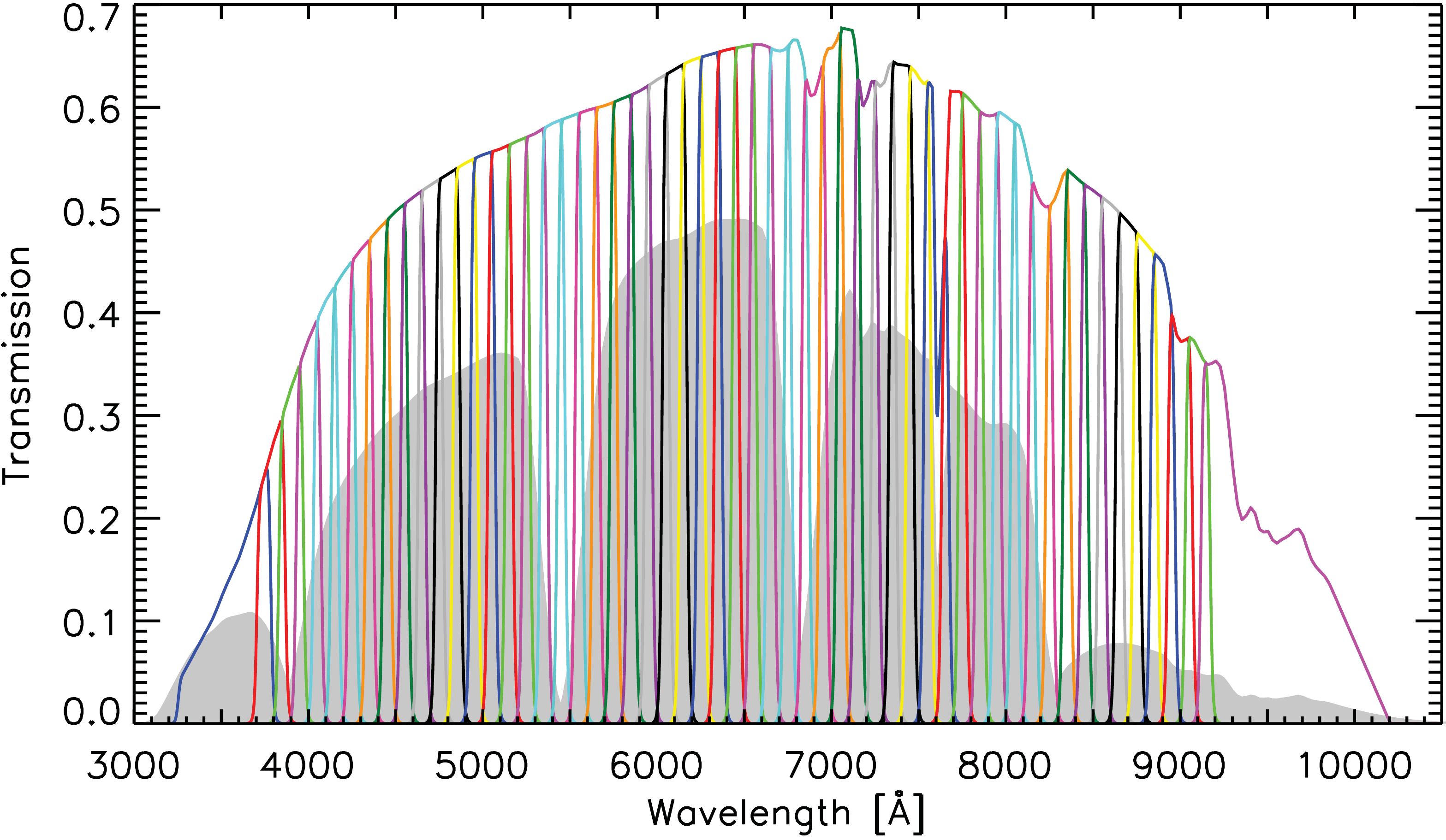}
\caption{
The transmission curves that characterize the quasi-spectroscopy of J-PAS.
Shown are the 54 narrow-band and 2 medium-band  filters that span the optical range. The narrow-band filters feature a width of $145$ \AA~and are spaced by 100 \AA. Also shown (gray areas) are the five SDSS filters. From \cite{Ascaso:2016ddl}.}
\label{fig:j-pas_filters}
\end{figure}
%%%%%%%%%%%%%%%%%%%%%%%%%%%%%%%%%%%%%%%%%%%%%%%%%%%%%

The Javalambre Physics of the Accelerating Universe Astrophysical Survey~\citep[\jpas --][]{Benitez:2014ibt} is a ground-based survey that is expected to begin scientific observations at the beginning of 2019. It features a dedicated 2.5m telescope with an excellent étendue which sports a 1.2 Gigapixel camera with a very large field of view of $4.7 \deg^2$. The observatory is in the mountain range ``Sierra de Javalambre'' (Spain), at an altitude of 2000 meters, an especially dark region with the very good median seeing of $0.7''$.

\jpas will observe approximately $8500 \deg^2$ of the sky via the revolutionary technique of quasi-spectroscopy: by observing with 54 narrow-band filters, two medium-band filters and three broad-band filters it will produce a pseudo-spectrum ($R\sim 50$) for every pixel, see Figure~\ref{fig:j-pas_filters}.
Therefore, \jpas really sits between photometric surveys such as DES and spectroscopic surveys such as DESI, fruitfully combining the advantages of the former (speed and low cost) with the ones of the latter (spectra).
In particular, it will be possible to determine the redshift of galaxies with a precision of $0.003(1+z)$.
In other words, it will be possible to accurately study the large scale structure of the universe using the galaxy and quasar catalogs produced by \jpas. This makes \jpas the first survey to approach the ``Stage IV'' level.

\begin{figure}
\centering %trim={<left> <lower> <right> <upper>}
\includegraphics[width= 9 cm]{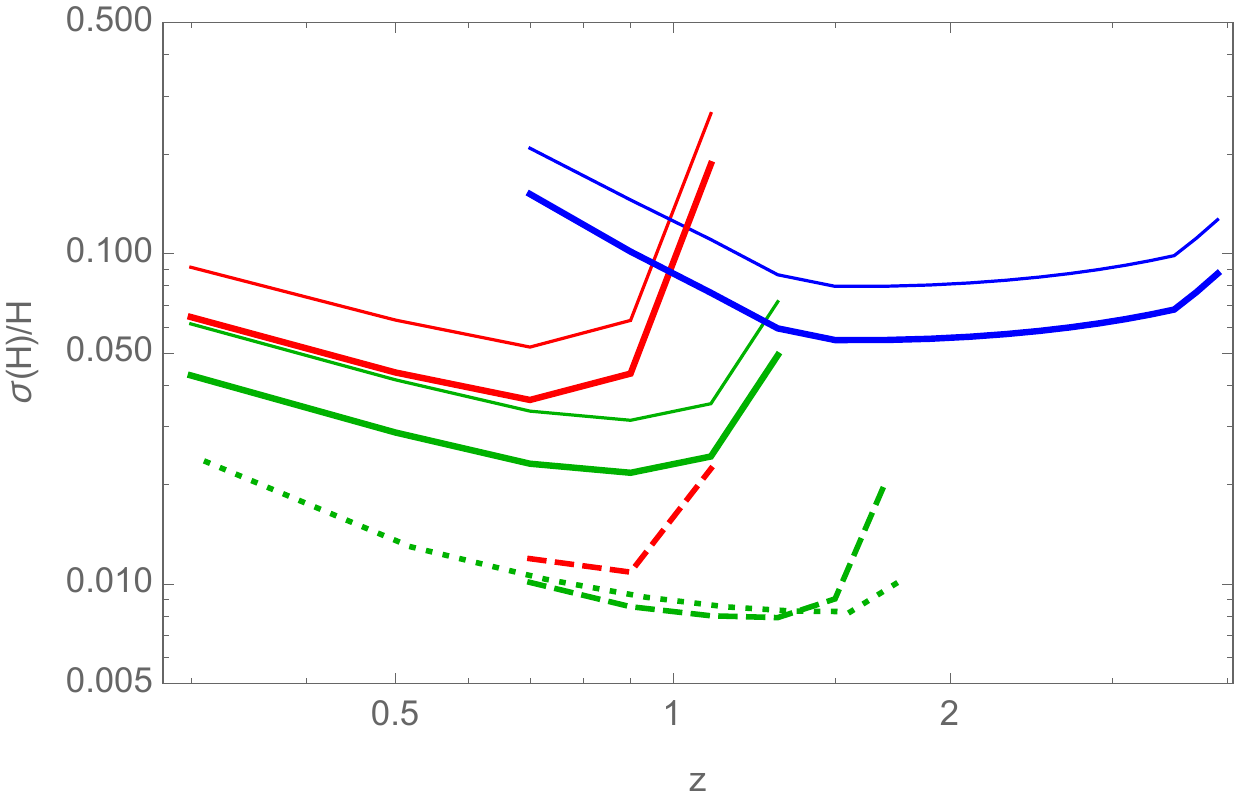}
\includegraphics[width=9 cm]{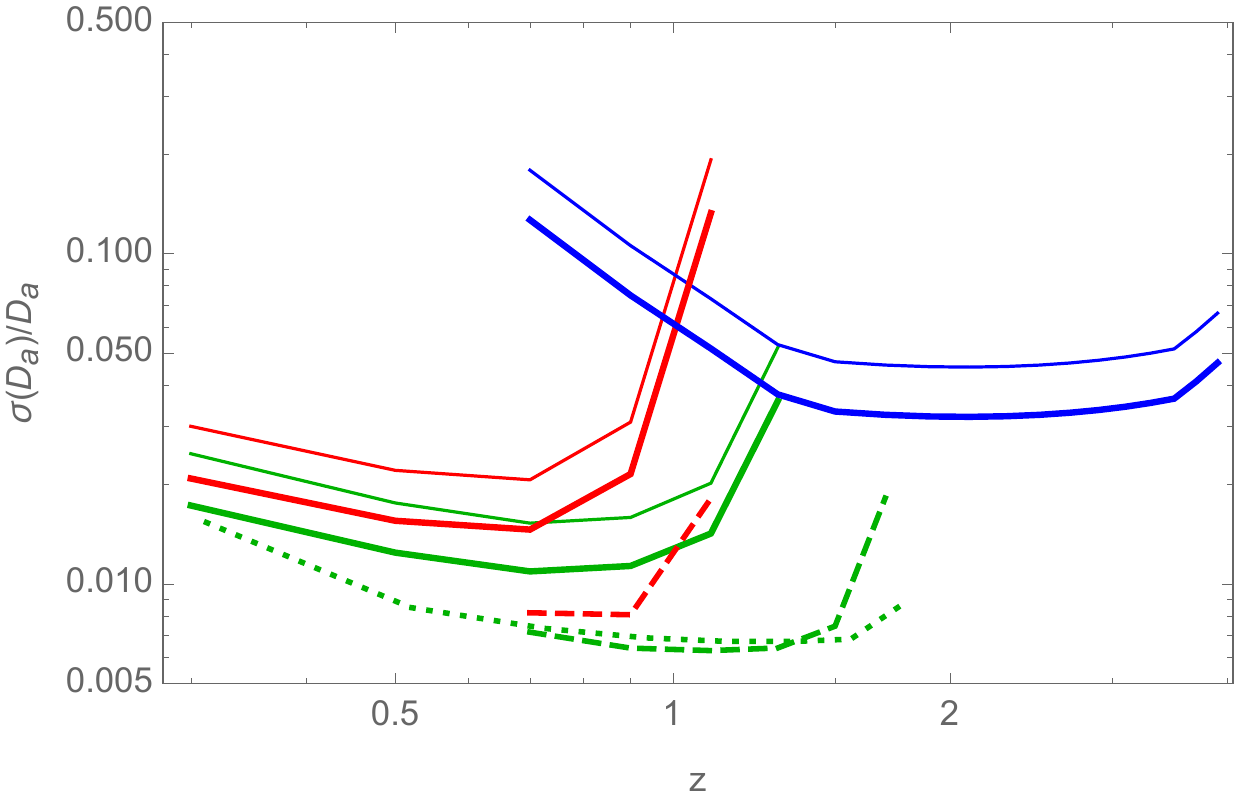}
\caption{Forecasted constraints on radial ($H$) and angular ($D_a$) BAO relative error from the galaxy catalogs produced by \jpas (thick solid lines for $8,500 \deg^2$, thin solid lines for $4,000 \deg^2$) as compared with DESI (dashed lines, $14,000 \deg^2$) and Euclid (dotted lines, $20,000 \deg^2$).
LRGs in red, ELGs in green and QSOs in blue. Redshift bins of $\Delta z = 0.2$ are used~\citep{forebramo}.}
\label{fig:j-pas_BAO}
\end{figure}

\begin{figure}
\centering %trim={<left> <lower> <right> <upper>}
\includegraphics[width=9 cm]{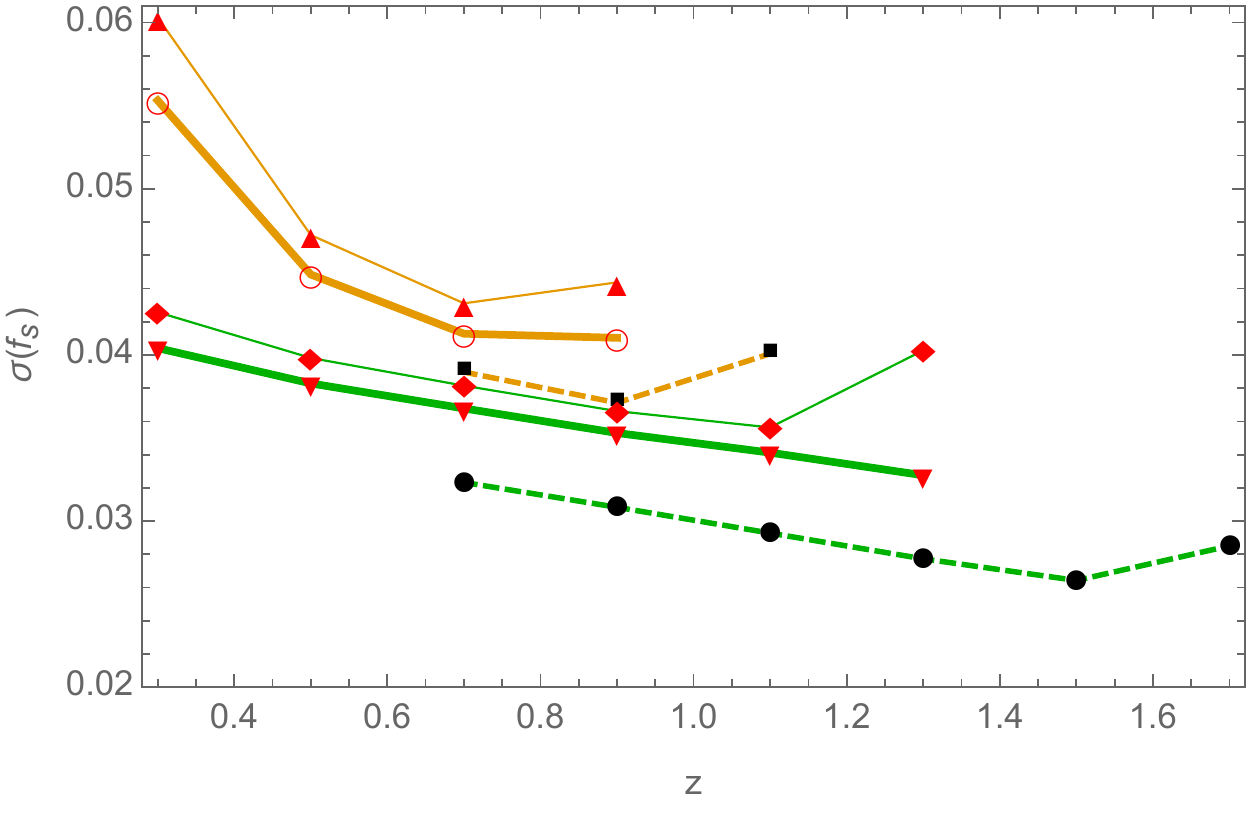}
\caption{Forecasted constraints on the growth of structure ($f_s= f \sigma_8$) relative error from the galaxy catalogs produced by \jpas (thick solid lines for $8,500 \deg^2$, thin solid lines for $4,000 \deg^2$) as compared with DESI (dashed lines, $14,000 \deg^2$).
LRGs in mustard and ELGs in green.
$f$ is the growth rate of matter perturbation.
Redshift bins of $\Delta z = 0.2$ are used~\citep{forebramo}.}
\label{fig:j-pas_fs8}
\end{figure}

As far as the dark sector and modified gravity, the most interesting observables will be galaxy clustering and galaxy cluster number counts. Regarding the former, thanks to the very precise photo-$z$ determinations and the large volume that will be explored, it will be possible to obtain excellent measurements of  Baryonic Acoustic Oscillations (BAO) and Redshift Space Distortions (RSD) in a wide redshift range ($0<z<3$).
About 90 million luminous red galaxies (LRG) and emission line galaxies (ELG) (up to $z\sim 1.2$) and 2 million quasars (up to $z\sim 3$) are expected to be detected.
Figures \ref{fig:j-pas_BAO} and \ref{fig:j-pas_fs8} show the corresponding forecasts. See also \cite{Abramo:2013awa,Abramo:2015iga} where constraints using the multi-tracer method are discussed.

%%%%%%%%%%%%%%%%%%%%%%%%%%%%%%%%%%%%%%%%%%%%%%%%%%%
%%%%%%%%%%%%%%%%%%%%%%%%%%%%%%%%%%%%%%%%%%%%%%%%%%%%%

%%%%%%%%%%%%%%%%%%%%%%%%%%%%%%%%%%%%%%%%%%%%%%%%%%%
%%%%%%%%%%%%%%%%%%%%%%%%%%%%%%%%%%%%%%%%%%%%%%%%%%%%%

\begin{figure}
\centering %trim={<left> <lower> <right> <upper>}
\includegraphics[width= 8 cm]{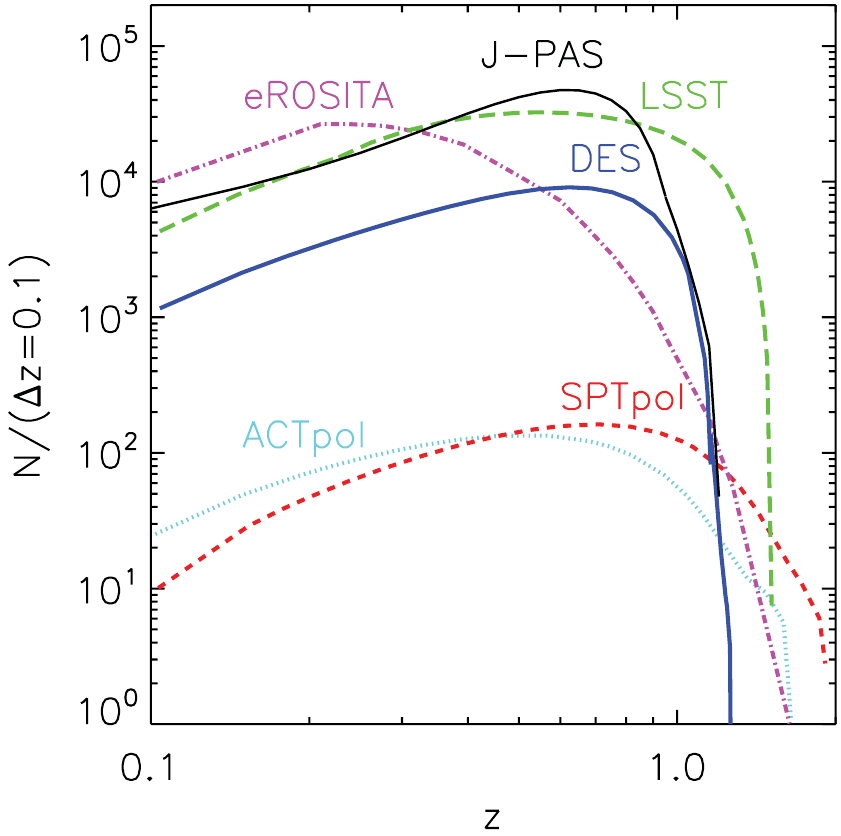}
\caption{Total number of groups/clusters per redshift bin as a function of redshift for different next-generation surveys. From \cite{Ascaso:2016ddl}.}
\label{fig:j-pas_clu}
\end{figure}

Regarding cluster counts, thanks again to its quasi-spectroscopic photometric redshift, \jpas will be able to separate cluster members from foreground and background galaxies with very high accuracy.
Indeed, the accuracy of the photometric redshift matches the typical velocity dispersion of massive clusters:
this ability together with the large area covered will allow J-PAS to detect clusters to much lower masses and higher redshifts than conventional photometric wide-field surveys.
\jpas will produce a catalog of about 700 thousand clusters with more than 10 members, down to $\sim 3 \cdot 10^{13} M_{\odot}$. See Figure \ref{fig:j-pas_clu} for a forecast.
Weak lensing observations will also be carried out and will be used to calibrate the cluster mass determination.

%%%%%%%%%%%%%%%%%%%%%%%%%%%%%%%%%%%%%%%%%%%%%%%%%%%
%%%%%%%%%%%%%%%%%%%%%%%%%%%%%%%%%%%%%%%%%%%%%%%%%%%%%

Weak and strong gravitational lensing data will also contain important cosmological information.
\jpas will be a revolutionary observatory also regarding the study of supernovas, galaxy evolution and stellar physics. See \cite{Benitez:2014ibt} for the full potential of the \jpas survey.

\subsection{Dark Energy Spectroscopic Instrument}

The Dark Energy Spectroscopic Instrument (DESI) is a Stage IV ground-based dark energy experiment that will study via a wide-area galaxy and quasar redshift survey both the expansion history of the universe through baryon acoustic oscillations and the growth of structure through redshift-space distortions.
DESI will be the successor to the BOSS survey and will complement imaging surveys such as DES and LSST.  
DESI will strongly constrain the nature of dark energy, theories of modified gravity, inflation, and will provide tight limits on the sum of neutrino masses.

DESI will obtain optical spectra for tens of millions of galaxies and quasars, constructing a revolutionary 3D map spanning the nearby universe to 11 billion light years.
This feat will be achieved using 5,000 pencil-size robots that will position the optical fibers that will catch the light from distant objects and transmit it to the spectrographs. 
The DESI Survey will be conducted on the Mayall 4-meter telescope at Kitt Peak National Observatory in Arizona (USA), starting at the beginning of 2020. 
See figures \ref{fig:j-pas_BAO} and \ref{fig:j-pas_fs8} for forecasts on radial and angular BAO and on the growth of structures.
See \cite{Aghamousa:2016zmz} for further information.

\subsection{Euclid Consortium}

The Euclid spacecraft \citep{Laureijs:2011gra} is currently under construction and scheduled for launch in the second half of 2021. During its mission, which will last at least 6 years, Euclid will observe approximately $\Omega_{\rm sky}=15000 \deg^2$ of the extra-galactic sky, which is about half of the total sky facing away from the Milky Way.

Euclid is the combination of two complementary probes. The 1.2-m Korsch telescope will feed, via a beam splitter, the visible band imager (VIS) and the near infrared spectrometer and photometer (NISP) instruments, in step-and-stare mode.
Thanks to this unique design it will be possible to produce, at the same time, 40 million spectroscopic redshifts in the range $0.5<z<2$ and 2 billion galaxy images with photo-$z$ in the redshift range $0<z<3$.
In other words, Euclid will allow us to study simultaneously the clustering (the potential $\Psi$) \textit{and} the lensing (the combination $\Psi-\Phi$) of galaxies. It will so constrain both the potential $\Psi$ and $\Phi$, thus factoring out possible survey-specific systematics which could degrade the results obtained from the combination of the two observables.
The potentials $\Psi$ and $\Phi$ encode the growth of scalar perturbations which is still poorly constrained and could signal physics beyond the standard model of cosmology such as modifications to general relativity at large cosmological scales.

Figures \ref{fig:j-pas_BAO} and \ref{fig:ska} show the forecasted error on radial and angular BAO determinations.
From Euclid alone it will be possible to obtain a FoM on the dark energy equation of state greater than 400 and constrain the growth of perturbations at the level of $\sigma_{\gamma}=0.01$, where $\gamma$ parametrizes deviation from the growth rate of matter perturbation in the $\Lambda$CDM model: $f(z)=\Omega_m(z)^{\gamma}$. For $\Lambda$CDM it is $\gamma \simeq 0.55$.
If Euclid data will be consistent with $\Lambda$CDM, this level of precision will allow us to confirm the standard model of cosmology with a ``decisive'' statistical evidence (using Jeffreys' scale terminology).

Also, it will be possible to identify 60 thousand clusters in the redshift range $0.5<z<2$, with more than 10 thousand at $z>1$. See the review \cite{Amendola:2016saw} for the full breadth of the Euclid mission.

\subsection{Large Synoptic Survey Telescope}

The Large Synoptic Survey Telescope (LSST)\footnote{\href{http://lsst.org}{lsst.org}.} is a  wide-field, ground-based, 8m-class telescope that is designed to image every few nights a substantial fraction of the sky in the six optical bands ugrizy (320-1050 nm). The 8.4-meter LSST uses a special three-mirror design to create an exceptionally wide field of view of 9.6 deg$^2$ (roughly 49 times the area of the Moon in a single exposure), and has the ability to survey the entire sky in only three nights. LSST will be equipped with the largest digital camera ever built, with 3.2 billions of pixels tiled by 189 4k x 4k CCDs.
The main survey will feature a homogeneous depth across approximately 20,000 deg$^2$ of sky, which will be covered with pairs of 15-second exposures in two photometric bands every three nights.
LSST aims at yielding high image quality and excellent astrometric and photometric accuracy. The coadded data will have the remarkable depth of $r \sim27.5$. LSST’s wide and deep coverage of billions of galaxies has the power to test differences in fundamental models that describe the Universe.

 The LSST is currently being built on the Cerro Pachón ridge at CTIO, Chile. Construction has started in 2014, first light is expected for 2019, Science Verification is scheduled for 2020 and Science Operations should start in 2023. The survey is planned to operate for a decade and is designed to meet the requirements of a diverse range of science goals in cosmology, astronomy and astrophysics, including the study of dark matter and dark energy. Much of that power comes from the fact that the measurements will be obtained from the same basic set of observations, using a powerful facility that is optimized for the purpose.

The Science case for the LSST is described in the LSST Science book \citep{Abell:2009aa}.\footnote{\href{http://lsst.org/scientists/scibook}{lsst.org/scientists/scibook}.}
In 2008, eleven separate science collaborations were formed in order to study the science that the LSST could carry out.
The one directly involved with the study of Dark Energy is the Dark Energy Science Collaboration (DESC). Within the DESC there are several working groups:
\begin{itemize}
\item Theory and Joint Probes,
\item Weak Lensing,
\item Large Scale Structure,
\item Supernovae,
\item Strong Lensing,
\item Photometric Redshifts.
\end{itemize}
The science goals regarding dark energy are:
\begin{itemize}

\item Weak gravitational lensing:
the bending/distortion of the light of distant sources from dark and baryonic matter along the line of sight.
Tomographic weak lensing measurements  will yield percent-level constraints on the nature of the dark sector and modified gravity.

\item Large-scale structure:
the vast number of galaxies that will be detected by LSST will allow us to measure the Baryonic Acoustic Oscillations and the distance-redshift relation with percent-level precision.

\item Type Ia Supernovae: LSST will discover tens of thousands of well-measured supernova light curves up to $z \sim 1$ over the full ten-year survey, yielding an accurate determination of the luminosity distance-redshift relation.

\item Galaxy clusters:
LSST will measure the masses of $\sim$20,000 clusters with a precision of $10\%$, which will give information about their distribution as a function of redshift.

\item Strong gravitational lensing:
LSST will produce a sample of $\sim$2600 time-delayed lensing systems, an increase of two orders of magnitude compared to present-day samples.
Angular-displacement, morphological-distortion and time-delay information will allow us to constrain the massive lensing objects. 

\end{itemize}

\begin{figure}
\centering
\includegraphics[width=7.5 cm]{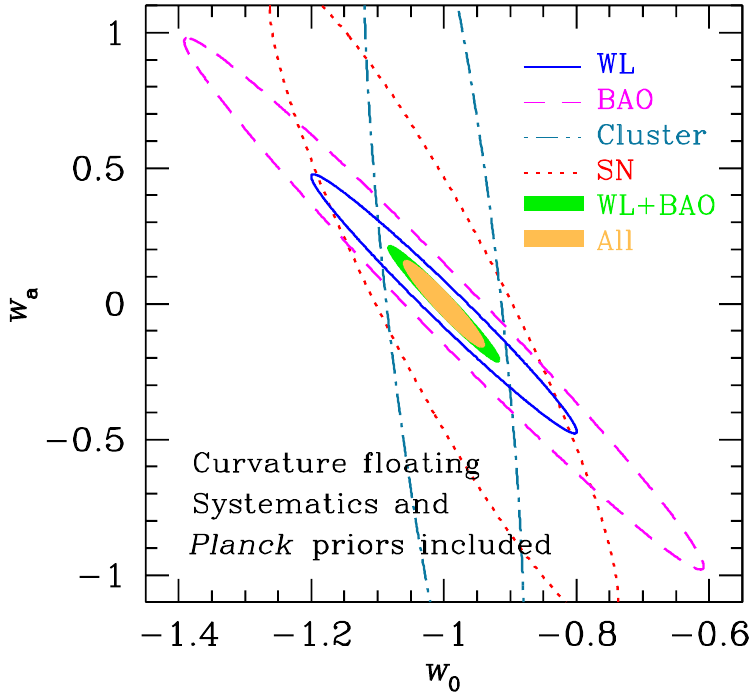}
\caption{1-$\sigma$ Fisher forecast for $w_0$ and $w_a$ from future LSST BAO (dashed line), cluster counting (dot-dashed line), supernovas Type Ia (SN, dotted line), Weak Lensing (WL, solid line), BAO + WL (green area), and all combined (yellow  area). The BAO and WL constraints are based only on galaxy-galaxy, galaxy-shear, and shear-shear power spectra.
From \citet{Abell:2009aa}.}
\label{fig:Fisher}
\end{figure}

LSST is a natural evolution of DES. Both are photometric surveys using digital cameras. DES is now finishing its 5.5 years of observations. However, the dark energy constraining power of LSST could be several orders of magnitude greater than that of the DES. In  Fig.~\ref{fig:Fisher} it is shown the Fisher matrix forecast for the LSST sensitivity on the parameters $w_0$ and $w_a$ of equation \eqref{weq}. It is clear the importance of combining different probes in order to obtain better constraints.

%Fisher LSST
%%%%%%%%%%%%%%%%%%%%%%%%%%%%%%%%%%%%%%%%%%%%%%%%%%%
%%%%%%%%%%%%%%%%%%%%%%%%%%%%%%%%%%%%%%%%%%%%%%%%%%%%%

LSST will strongly test theories of modified gravity by accurately measuring the growth of structure.
However, it is worth stressing that the sheer statistical power of the LSST dataset will allow for unprecedented modeling of systematics as a variety of null tests and a multitude of nuisance parameters will be included in the analysis.
Furthermore, such a large and homogeneous catalog will allow for joint analysis which mitigates systematics and improve calibration.
For example, instead of obtaining constraints on dark energy from cosmic shear and galaxy cluster counts separately, LSST may use clusters and galaxy-galaxy lensing simultaneously to reduce photometric redshift  and mass calibration errors.

%%%%%%%%%%%%%%%%%%%%%%%%%%%%%%%%%%%%%%%%%%%%%%%%%%%%%%%%%%%%%%%%
\section{Square Kilometer Array}

%%%%%%%%%%%%%%%%%%%%%%%%%%%%%%%%%%%%%%%%%%%%%%%%%%%
\begin{figure}
\centering %trim={<left> <lower> <right> <upper>}
\includegraphics[width= 8 cm]{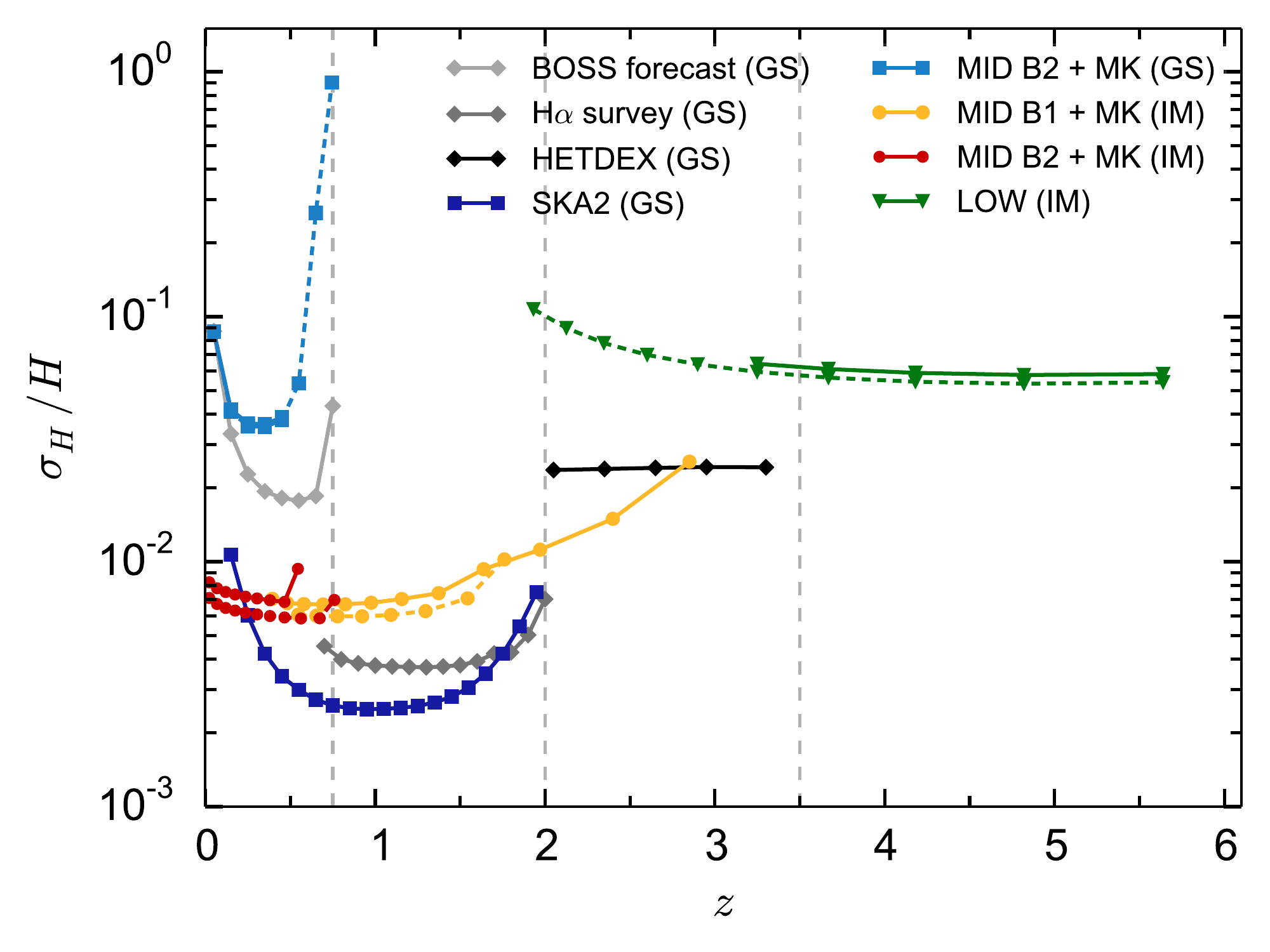}
\includegraphics[width=8 cm]{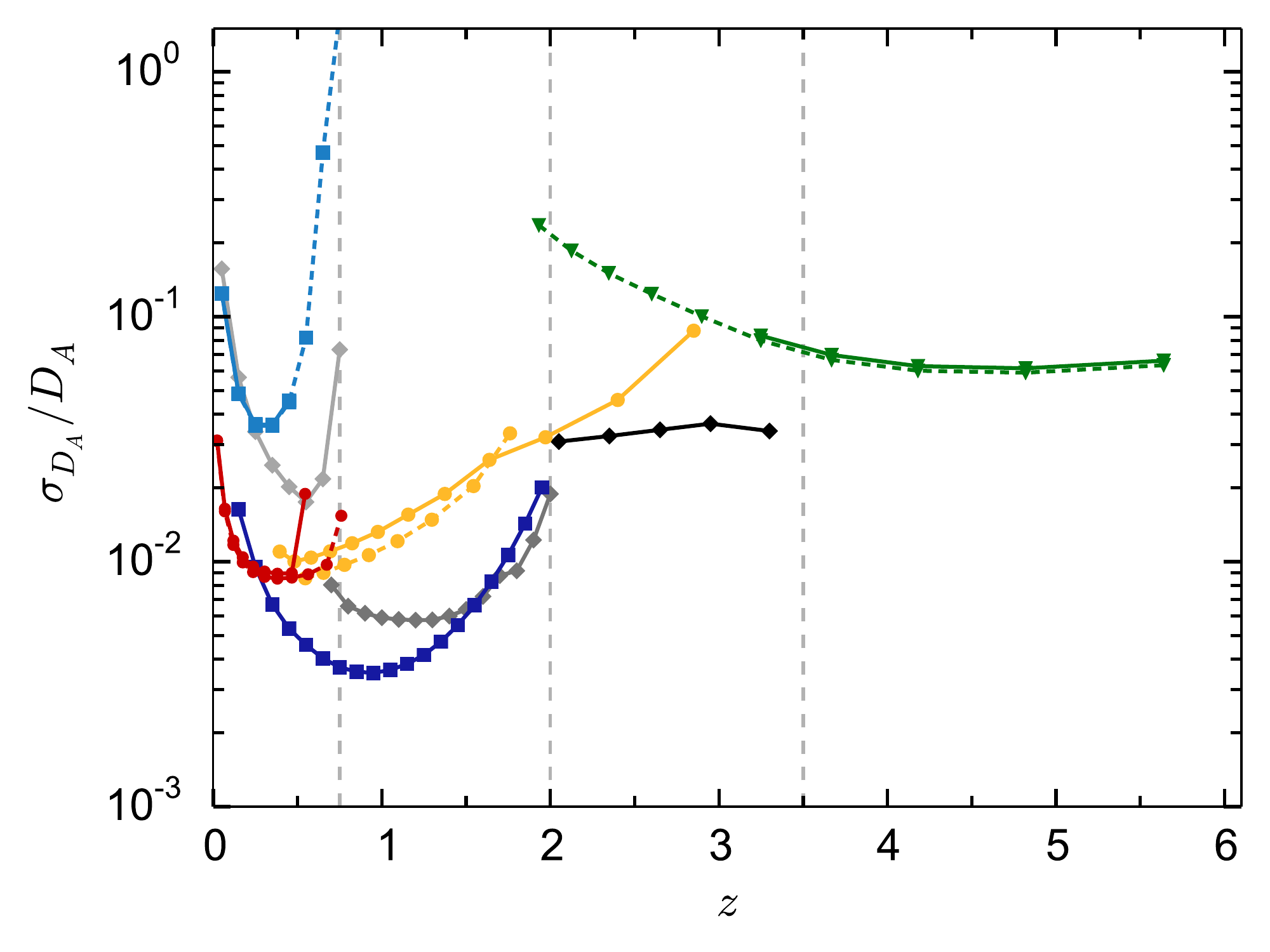}
\includegraphics[width=8 cm]{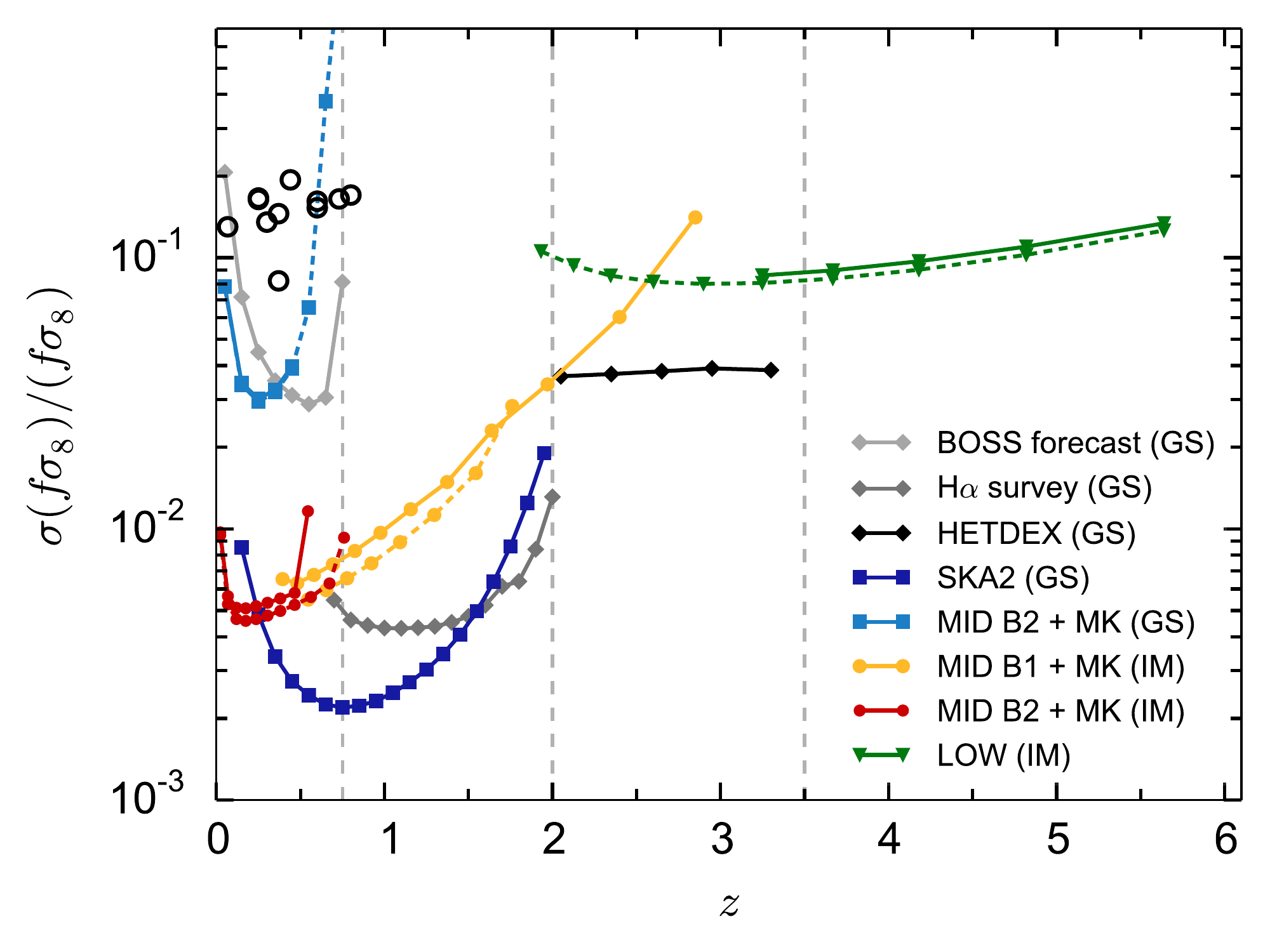}
\caption{Forecasted constraints on radial ($H$) and angular ($D_A$) BAO and the growth of structures ($f \sigma_8$) from SKA as compared with other surveys. ``GS'' stands for galaxy survey while ``IM'' for intensity mapping survey (H$\alpha$ survey is a Euclid-like survey).
Forecasts from \cite{Bull:2015lja} where more information can be found.}
\label{fig:ska}
\end{figure}
%%%%%%%%%%%%%%%%%%%%%%%%%%%%%%%%%%%%%%%%%%%%%%%%%%%%%

Another revolutionary future survey is the Square Kilometer Array (SKA), which will become the world's largest radio telescope, featuring a total collecting area of approximately one square kilometer.
It will operate over a wide range of frequencies and its size will make it 50 times more sensitive than any other radio instrument. 
It will be built in two phases.
Phase 1 is expected to end observations in 2023 and will be split into SKA1-SUR (Australia) and SKA1-MID (South Africa).
Phase 2 is scheduled for 2030 and will be at least 10 times as sensitive \citep[see][]{Yahya:2014yva, Santos:2015hra, Raccanelli:2015hsa, Bull:2015nra}.

The SKA will survey the large-scale structure by detecting the redshifted neutral hydrogen 21cm emission line from a large number of galaxies out to high redshift. This can be achieved in two ways: by measuring the 21cm line for many individually-detected galaxies (a galaxy redshift survey) or by measuring the large-scale fluctuations of the integrated 21cm intensity from many unresolved galaxies (intensity mapping). The SKA surveys will cover a combined survey volume and redshift range that is significantly larger than that of even Euclid and LSST.

SKA1 will measure, in a sky area of 5000 $\text{deg}^2$  and a redshift range $z\le0.8$, approximately 5 million galaxies; SKA2 is expected to observe 30000 $\text{deg}^2$, reaching much higher redshifts ($z\le2.5$), and to detect approximately 1 billion  galaxies with spectroscopic redshifts \citep{Santos:2015hra}.
See Figure~\ref{fig:ska} for the forecasted constraints on radial and angular BAO and the growth of structures from SKA as compared with other surveys.

The SKA survey will allow us to address important questions on fundamental physics, in areas such as cosmic dawn and reionization, gravity and gravitational radiation, dark energy and dark matter, and astroparticle physics.
SKA will also shed light on the nature of neutrinos, cosmic inflation (early universe) and foundations of cosmology.  
See \citep[][and references therein]{Bull:2018lat} for a review of the fundamental physics that can be studied with the Square Kilometer Array.

%%%%%%%%%%%%%%%%%%%%%%%%%%%%%%%%%%%%%%%%%%%%%%%%%%%%%%%%%%%%%%%%
\section{Gravitational wave surveys}

The detection of GW170817 \citep{TheLIGOScientific:2017qsa}, the coincident
Gamma Ray Burst (GRB) \citep{Monitor:2017mdv}, and the other electro-magnetic
counterparts in a wide region of the spectrum from X to radio frequencies
\citep{GBM:2017lvd} marked the historical debut of Gravitational Waves (GWs)
on the stage of Multi-messenger Astronomy in the first month of joint activity of
the Advanced LIGO \citep{Harry:2010zz}, located in the US, and Advanced Virgo detector
\citep{TheVirgo:2014hva}, located in Italy.

Advanced LIGO and Advanced Virgo GW detectors are Michelson interferometer with
Fabry-Perot cavities which represent
the most precise ruler ever made: by measuring the differential variation of
the interferometer's arms they can monitor the passage of a GW in the frequency
range from few tens of Hz to roughly $1$ kHz. Because of the frequency range,
interferometric GW detectors
are sensitive only to binary coalescence of \emph{compact} objects, thus
small enough ($\sim 10-100$ km) that can achieve such high orbital frequencies.
Interferometers respond linearly to the GW strain by measuring the difference in
optical path with the result of being mild directional detectors, as they can
detect only GWs that do not alter symmetrically the two end mirrors.

The cryogenic Japanese detector KAGRA \citep{Somiya:2011np,Aso:2013eba},
with comparable design sensitivity, is planning to join the GW 
detection effort before the end of third 
Observation Run (O3) of LIGO and Virgo, which is due to start in April 2019
and to last for at least one year, and the Indian INDIGO \citep{Indigo}
by the start of the next decade.

\begin{figure}%[htb]
\centering
\includegraphics[width=.4\linewidth]{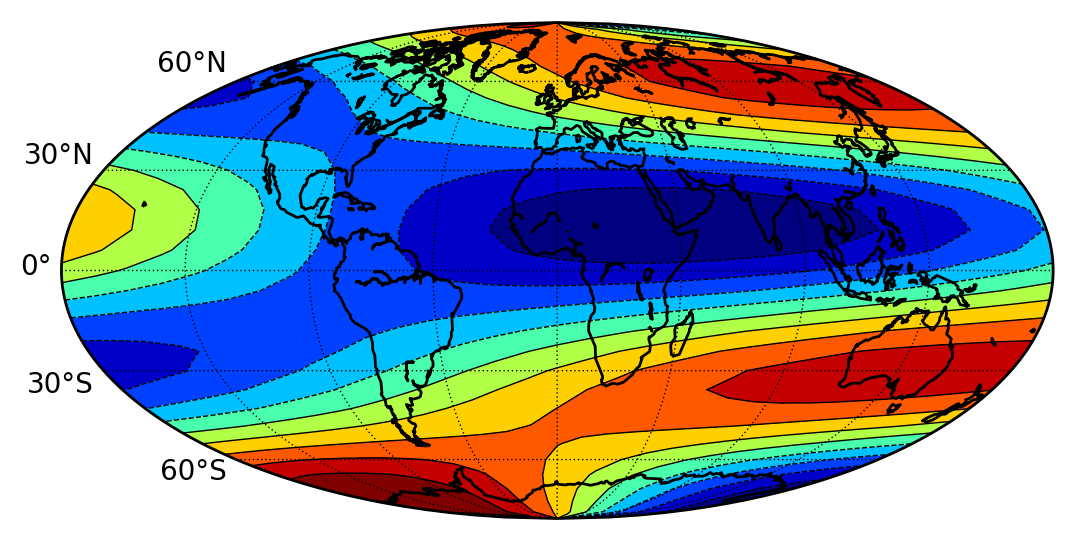}
\includegraphics[width=.4\linewidth]{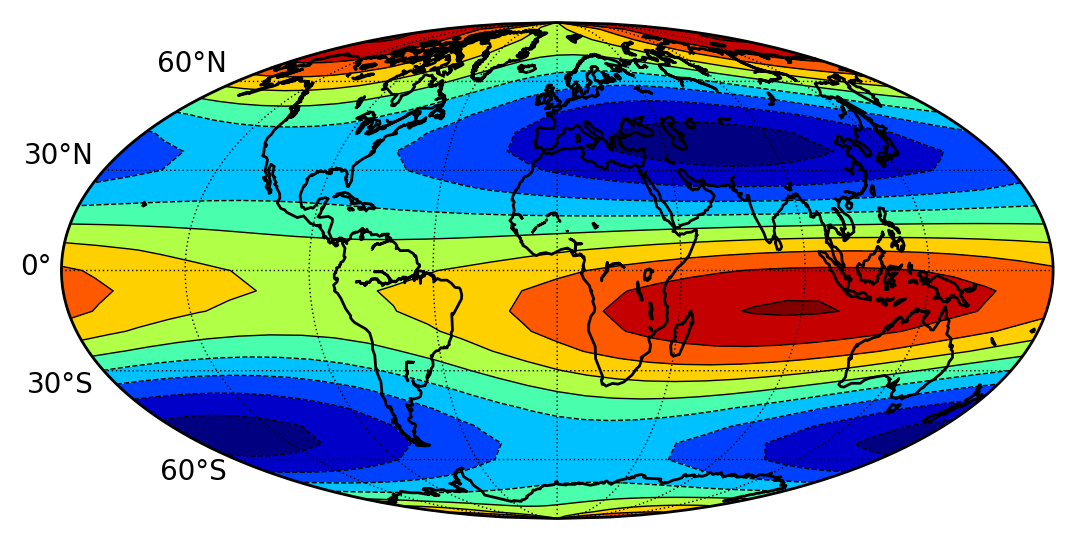}\\
\includegraphics[width=.4\linewidth]{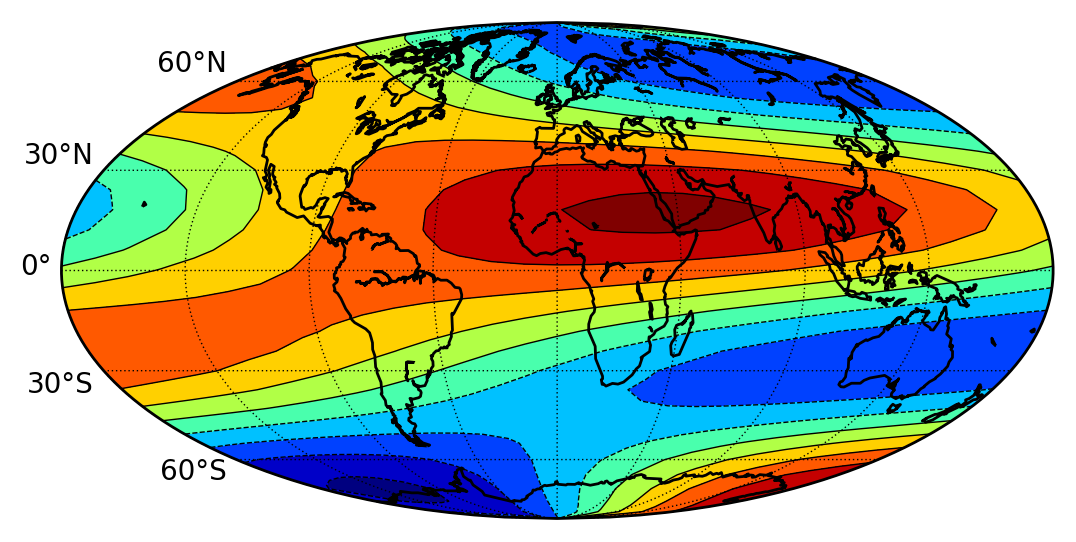}
\includegraphics[width=.4\linewidth]{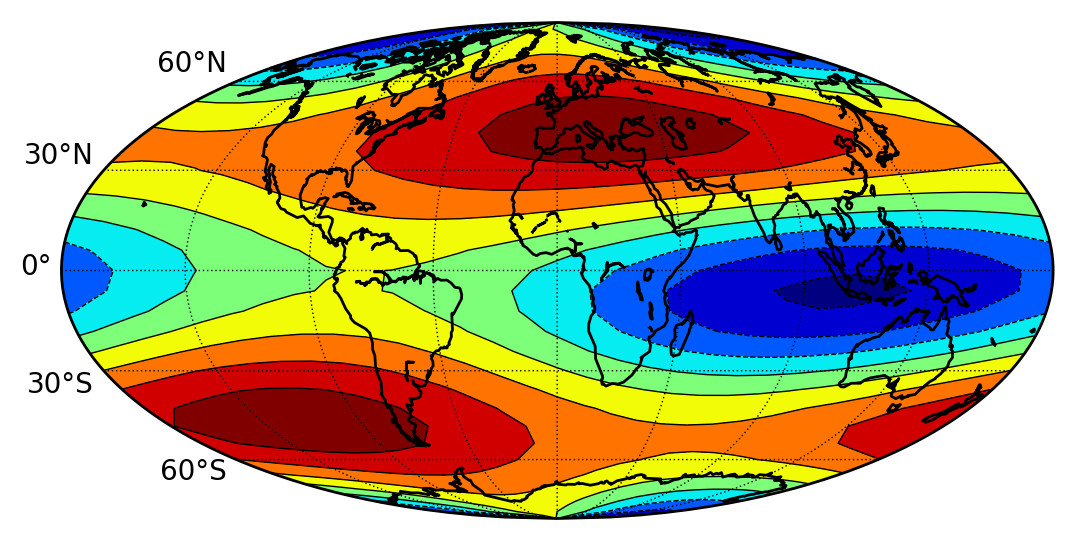}\\
\includegraphics[width=.4\linewidth]{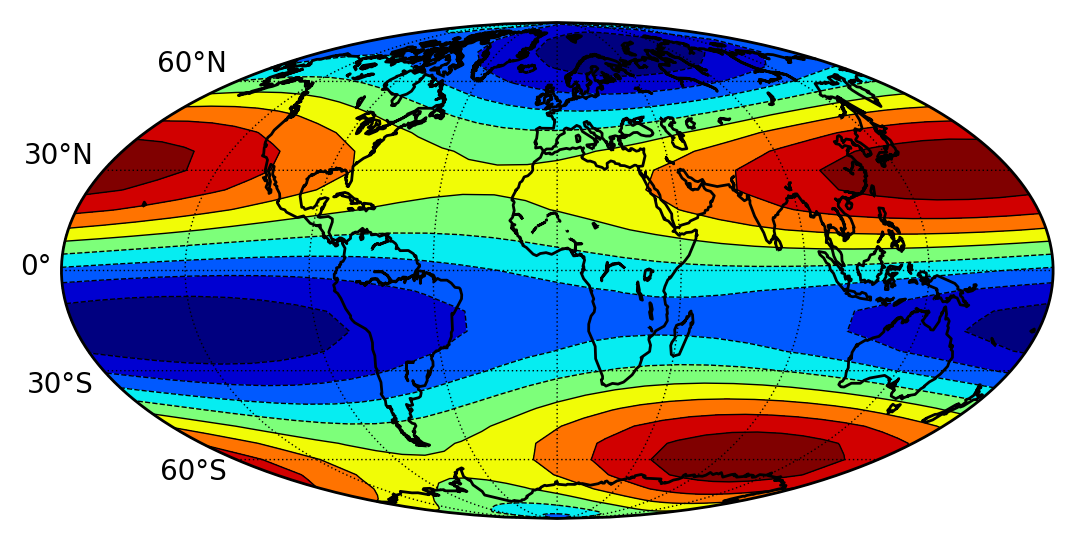}
\includegraphics[width=.4\linewidth]{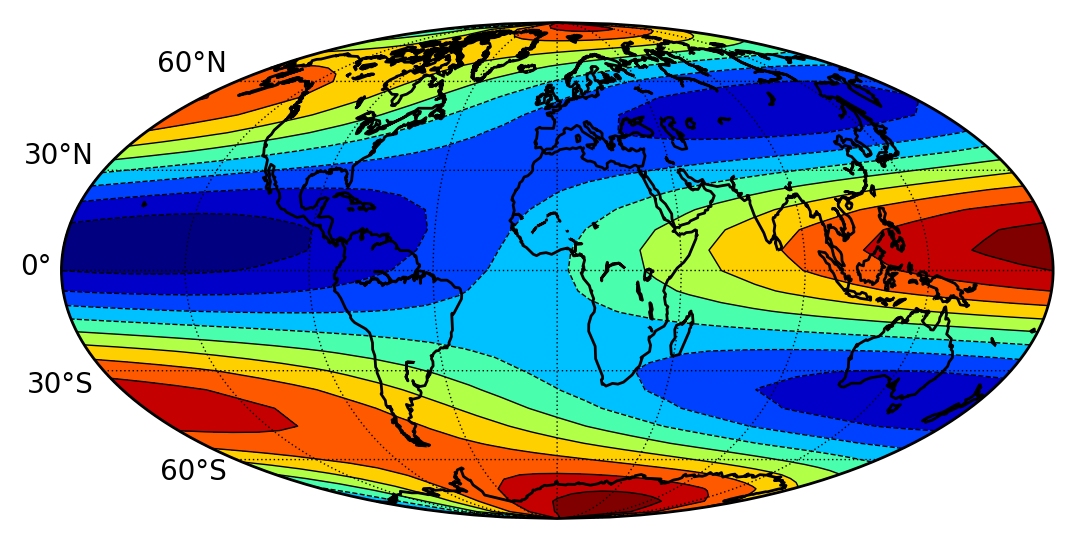}
\caption{Pattern functions of the LIGO Hanford (first line), LIGO Livingstone
(second line) and Virgo detector (third line) as a function of right ascension
and declination at the time of GW170817: 17 August 2017, 12:41:53 UTC. The first and second column represent respectively
$F_+$ and $F_\times$, the position of the GW170817 source being right ascension$=13h\,09'\,48''$, declination$=-23^o\,22'\,53''$.
Pattern function values range from 1 (dark red) to -1 (dark blue).
The values of $\sqrt{F_+^2+F_\times^2}$ for LIGO Hanford,LIGO Livingstone and
Virgo are respectively 0.89, 0.75, 0.30 at the location and time of GW170817. Computed via the LALSuite library \citep{lalsuite}.}
\label{fig:patterns}
\end{figure}

GWs have 2 polarizations, conventionally called $h_+$ and $h_\times$ and
each detector is sensitive to only one linear combination of them, the
coefficients of proportionality between detector output and $h_{+,\times}$ being
the \emph{pattern functions} $F_{+,\times}$, see Figure~\ref{fig:patterns} for the values
of the LIGO and Virgo pattern functions at the time of GW170817.
Note that LIGO is composed of 2 detectors and they are almost aligned, to have similar pattern functions so
no  event that is detected by one of the two can fall into the blind region of
the other.

For un-modeled events, LIGO and Virgo search for excess noise but for coalescing
binaries accurate theoretical models exist enabling to correlate observational
data with pre-computed templates.

One important quantitative detail is that because of the
quadrupolar nature of the source the two polarizations are affected in a
specific way by the relative orientation of the binary orbital plane and the
observation direction. Denoting such angle by $\iota$ one has
\begin{equation}
\begin{array}{rcl}
h_+&\propto& (1+\cos^2\iota)/2\,,\\
h_\times&\propto& \cos\iota\,,
\end{array}
\end{equation}
introducing a degeneracy between $\iota$ and the source-observer distance
to which the GW amplitude is inversely proportional: unless the two polarizations are independently measured there is a strong degeneracy between
distance and inclination.
Stronger signals could equally well be closer and misaligned or farther and better aligned, with 
the latter possibility favored a priori because at a larger
distance more volume is available, hence more sources are possibly
present
(until a redshift $z\sim 2$, see discussion below and \citet{Schutz:2011tw}).

\begin{figure}%[htb]
\centering
\includegraphics[width=.75\linewidth]{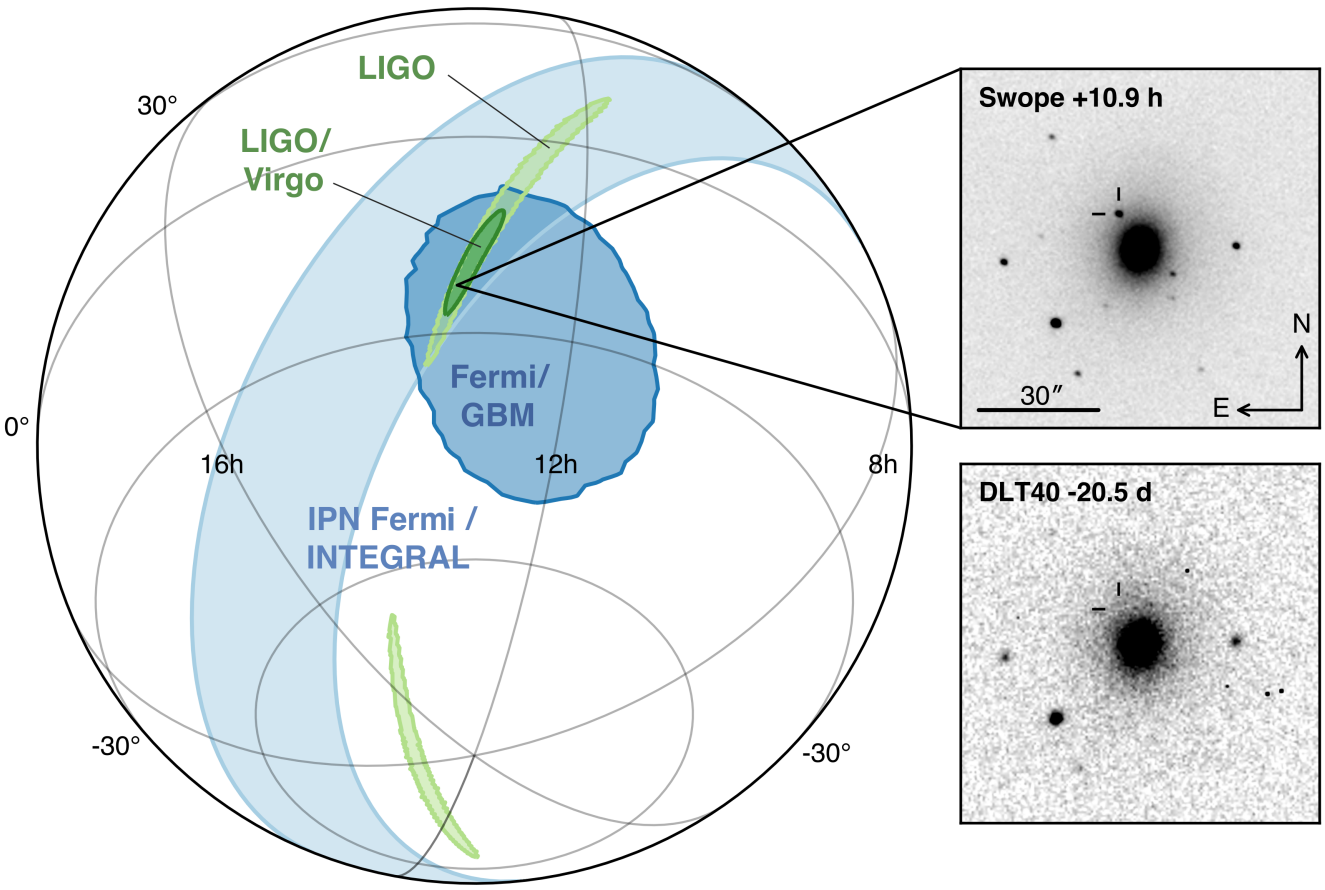}
\caption{Localization of the GW, gamma-ray, and optical signals. The left panel shows the 90\% credible regions from LIGO (light green),  LIGO-Virgo  (dark green), Fermi-INTEGRAL (light blue), and Fermi-GBM (dark blue). The inset shows the location of the  host galaxy NGC 4993 at 10.9 hr after the merger (top right) and from 20.5 days prior to merger (bottom right). The reticle marks the position of the transient in both images. From \citet{GBM:2017lvd}.}
\label{fig:local}
\end{figure}

GWs can be localized with reasonable accuracy, e.g., the 90\% credible region
of GW170817 which happened at 40 Mpc from Earth ($z \sim 0.01$) and was observed by 3
detectors (though very little signal was present in Virgo), measured $28$
degree squared, with lower precision expected for
fainter objects. The localization is obtained by short-circuiting the information
of the time of arrival (triangulation) and the information from the signal
amplitudes and phases across the detector network \citep{Aasi:2013wya}, with
the result shown in Figure~\ref{fig:local} for GW170817, where the GRB
\citep{GBM:2017lvd} and optical \citep{Coulter:2017wya} localizations are also
shown.

The almost coincident detection of GWs and GRB also enabled to constrain the
velocity of light and of GWs to be almost exactly equal to each other, up
to one part in $10^{-15}$ \citep{Monitor:2017mdv}, setting non-trivial constraint on
practically all non-General Relativity gravity model modifying the radiative
sector of General Relativity \citep{Creminelli:2017sry}.

\begin{figure}[t]
  \begin{center}
    \includegraphics[width=0.75\linewidth]{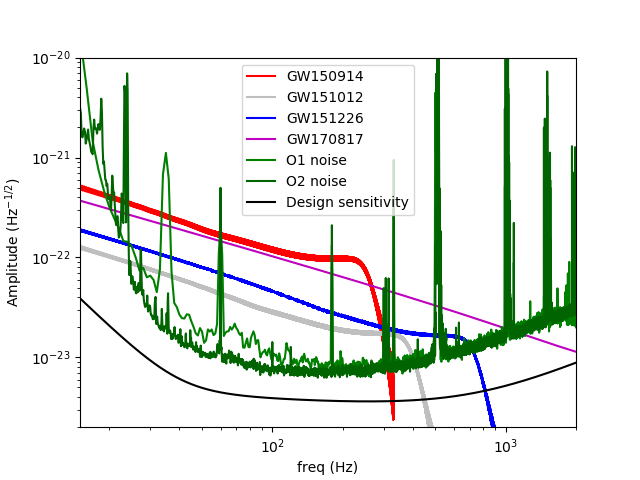}
    \caption{Spectrum of the 3 detected gravitational wave events in O1 and
      of GW170817
      compared to the real O1 and O2 noise (of the LIGO Livingstone detector)
      and to the Advanced LIGO design sensitivity. Data from the LIGO Open Science Center \citep{Vallisneri:2014vxa}.}
    \label{fig:sensO1Adv}
  \end{center}
\end{figure}

On the top of the GW event sourced by a binary neutron star, 10 more events have
been detected, 3 in the first Observation run O1 (lasted from  September 2015 to January 2016) and the remaining ones in O2 (spanning the period between December 2016 and August 2017, only the last month of which with both LIGOs and Virgo on), see Figure~\ref{fig:sensO1Adv} \citep{Abbott:2016blz,Abbott:2016nmj,Abbott:2017vtc,Abbott:2017gyy,Abbott:2017oio,LIGOScientific:2018mvr}.

The events detected are compatible with an event rate of $\sim 100$ merger
events per Gpc${}^{3}$ per year for binary black holes \citep{TheLIGOScientific:2016pea} and
$\sim 10^4$ merger events per Gpc${}^3$ per year for binary neutron stars
\citep{TheLIGOScientific:2017qsa}.
For comparison, the average density of galaxies is $\sim 10^8/$Gpc${}^3$.
With a distance reach, at design sensitivity, of $\sim 200$ Mpc for binary neutron stars,
and few Gpc for a black hole binary with a total mass of $\sim 100 M_\odot$,
one can realistically infer that up to one event per week will be detected in O3.

\begin{figure}
  \begin{center}
    \includegraphics[width=.75\linewidth]{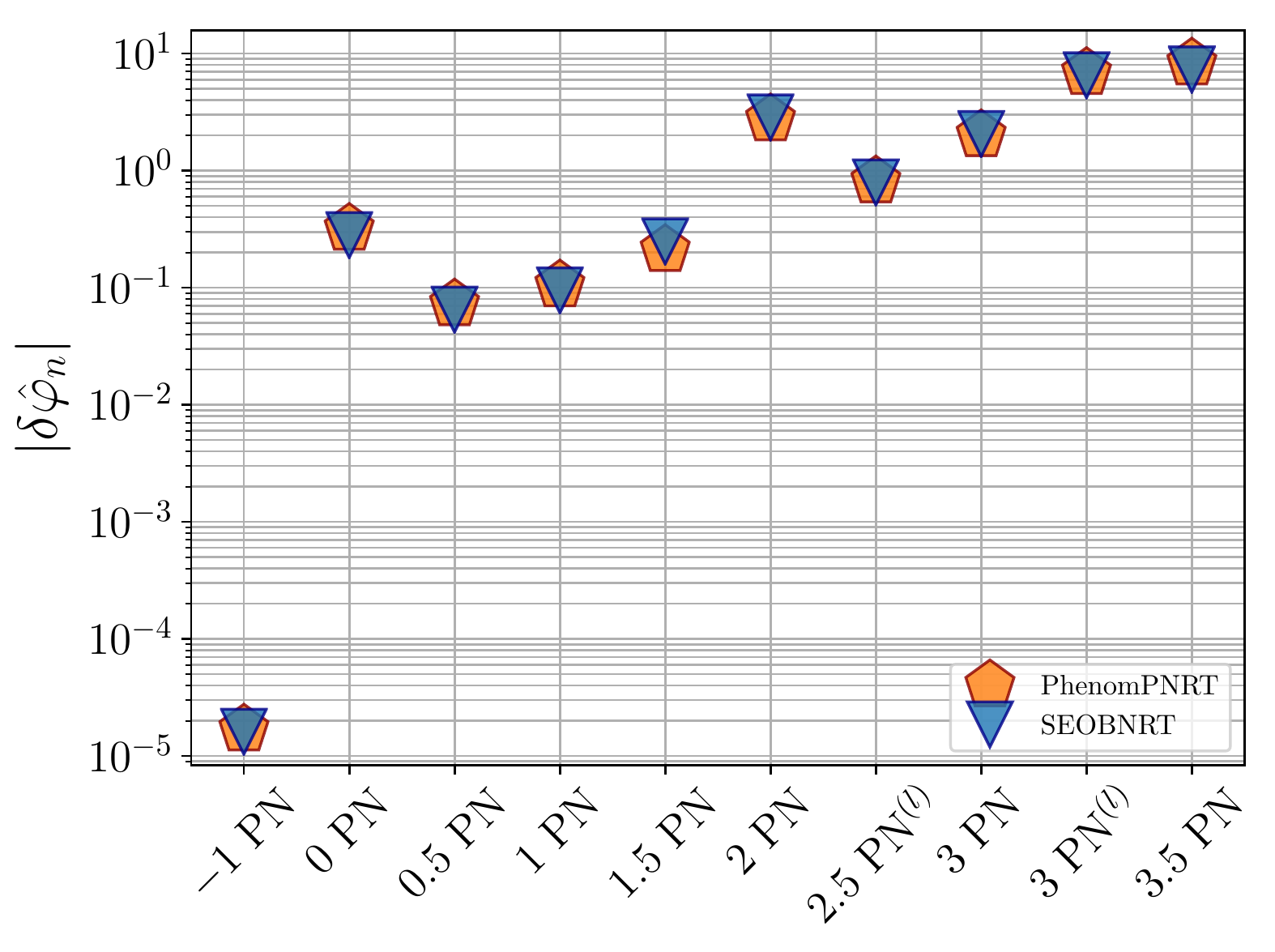}
    \caption{Bounds on deviation from phasing post-Newtonian coefficients from the analysis of the GW170817 signal. Note that the $-1$ and the $0.5$PN coefficients are identically zero in GR. Results for two
different phenomenological approximants IMRPhenomP \citep{Husa:2015iqa} and SEOBNR
\citep{Bohe:2016gbl} are reported. Different approximants are obtained by resumming the PN approximation
in different ways. From \citet{Abbott:2018lct}.}
    \label{fig:PNbounds}
  \end{center}
\end{figure}

On the fundamental physics side GW detections enabled the first ever constraint
on high order post-Newtonian parameters describing the 2 body dynamics.
The frequency $f$ of a signal changes as the binary distance shrinks and, at leading order, the rate of change of $f$ is given by
\be
\label{eq:fdot}
\dot f=\frac{96}5\pi^{8/3}\pa{G_NM_c}^{5/3}f^{11/3}\simeq 10 {\rm sec}^{-2}
\pa{\frac{M_c}{M\odot}}^{5/3}\pa{\frac f{100 {\rm Hz}}}^{11/3}\,,
\ee
where we have introduced the \emph{chirp mass} $M_c\equiv \eta^{3/5}M$, with
$\eta\equiv m_1m_2/M^2$, being $m_i$ the indivdual constituent mass and
$M\equiv m_1+m_2$.
It is possible to parametrize the observed GW phase $\phi$ in an expansion in terms of the relativistic parameter $v\equiv(G_NMf)^{1/3}$, being $G_N$
the Newton's constant:
\be
\phi(t)=\frac 5{16\eta} \int_{v_0}^{v(t)}\pa{1+\phi_1 v^2+\ldots+\phi_3{v^6}+\ldots}
\frac{dv}{v^6}\,,
\ee
where both fundamental gravity theory and astrophysical parameters of the
source concur to determine the \emph{post-Newtonian} coefficients $\phi_i$.
The most recent bounds are reported in  \cite{Abbott:2018lct}, see Figure~\ref{fig:PNbounds} relative to GW170817.

\begin{figure}
  \begin{center}
    \includegraphics[width=.8\linewidth]{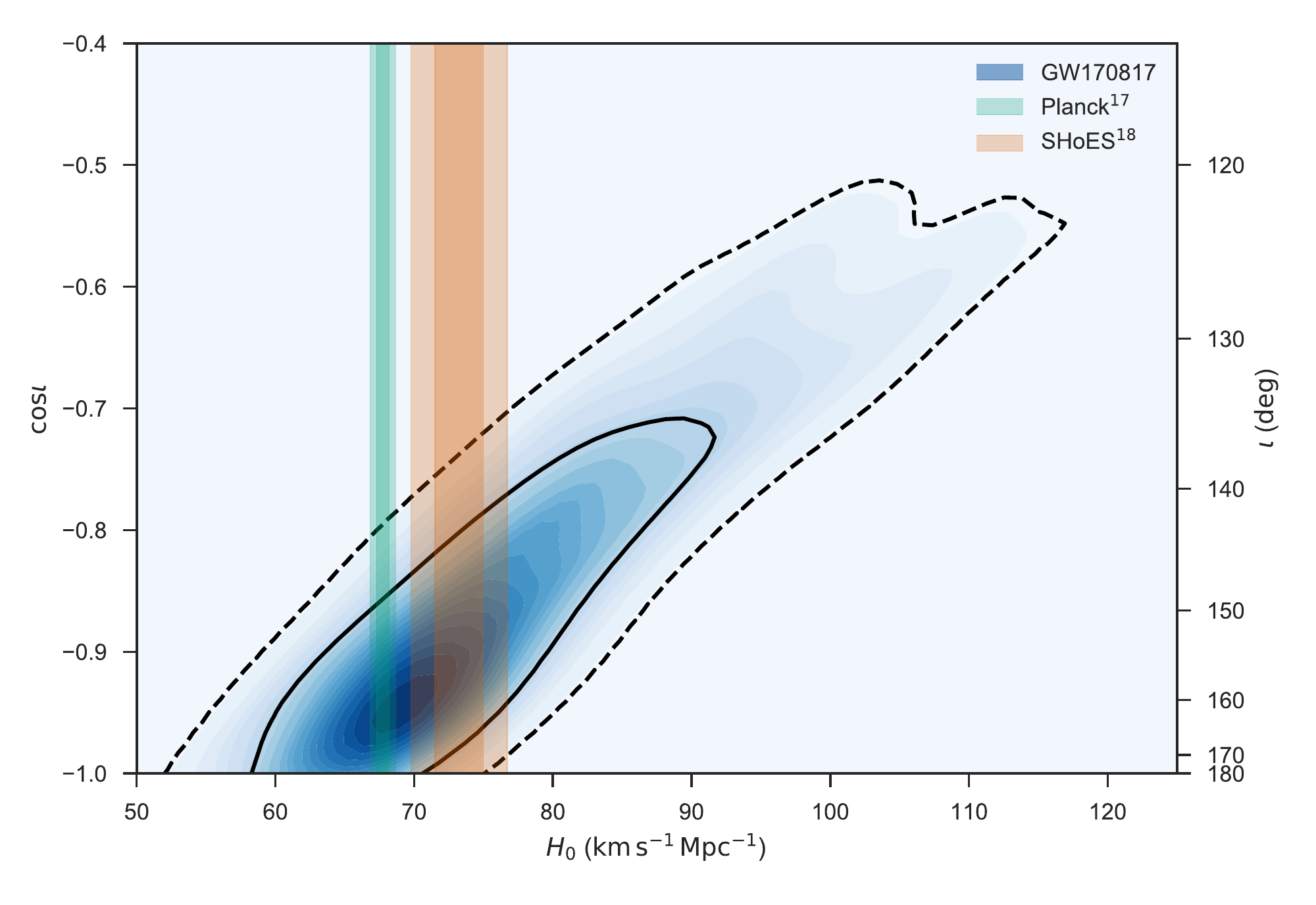}
    \caption{Two-dimensional probability distribution function of $\cos\iota$ and
      $H_0$ for the GW170817 event. Reported also the Hubble
      constant determination from Cepheid variable stars \citep{Riess:2016jrr} and CMB Planck data \citep{Ade:2015xua}.
      From \citet{Abbott:2017xzu}.}
    \label{fig:H0-iota}
  \end{center}

\end{figure}

On the cosmology side the coincident measure of luminosity distance via GWs and
redshift via electromagnetic radiation enabled the measure of the Hubble-Lema\^itre
constant, but with the nuisance of the correlation of luminosity distance with
the un-measured inclination angle $\iota$, giving the result in Figure~\ref{fig:H0-iota}.

Note that the GW signal does not allow to determine the redshift, since it is degenerate with the total mass of the binary. E.g.~in the phase $\phi(t)$ the
main dependencies are on the individual masses via the combination
$\phi(t_s/M_c,\eta)$ (it has additional, sub-leading dependence on the
dimension-less spins $\vec{\chi_{1,2}}\vec{S}_{1,2}/m^2_{1,2}$ and orbital angular momentum unit vector $\hat L$),
but substituting the source time $t_s$ for the observer time $t_o$ one gets
$\phi(t_o/((1+z)M_c),\eta)$, thus introducing the dependence on the 
the redshifted mass ${\cal M}\equiv M(1+z)$. E.g.~for the $+$ polarization,
denoting by $D$ the coordinate distance, we have
\be
\ba{rcl}
h_+ &=&\ds \frac{1+\cos^2\iota}2\, \eta
\frac {Mv^2}D\cos\phi\pa{t_s/M_c,\eta,\vec{\chi}_i\cdot\hat L/m_i^2,\vec{\chi}_1\cdot\vec{\chi}_2,\ldots}\\
&=&\ds  \frac{1+\cos^2\iota}2\, \eta\frac {M(1+z)v^2}{D(1+z)}
\cos\phi\pa{t_o/(M_c(1+z)),\eta,\vec{\chi}_i\cdot\hat L/m_i^2,\vec{\chi}_1\cdot\vec{\chi}_2,\ldots}\\
&=&\ds\frac{1+\cos\iota^2}2\eta
\frac {\mathcal{M}v^2}{d_L}\cos\left[\phi\pa{t_o/\mathcal{M},\eta,\vec{\chi}_i\cdot\hat L/m_i^2,\vec{\chi}_1\cdot\vec{\chi}_2,\ldots}\right]\,,
\ea
\ee
where the final result is expressed in terms of the \emph{luminosity distance}
$d_L=(1+z)D$.
The cross polarization has a similar expression, with a different pre-factor,
hence, beside not being able to disentangle $M$ and $z$ dependence, with
only one measurement of $F_+h_++F_\times h_\times$ it is also impossible to disentangle
$d_L$ and $\iota$, see Figure~\ref{fig:H0-iota}.

Redshift can be either measured electromagnetically or inferred from the luminosity distance \emph{assuming}
a cosmological model, in the latter case at the price of not being able
to \emph{check} the cosmological model.
GW170817 represented the first \emph{standard siren} event with electromagnetic
counterpart, and many more are expected in O3 at desgin sensitivity: $\sim O(1)/$month.

\begin{figure}
  \begin{center}
    \includegraphics[width=.75\linewidth]{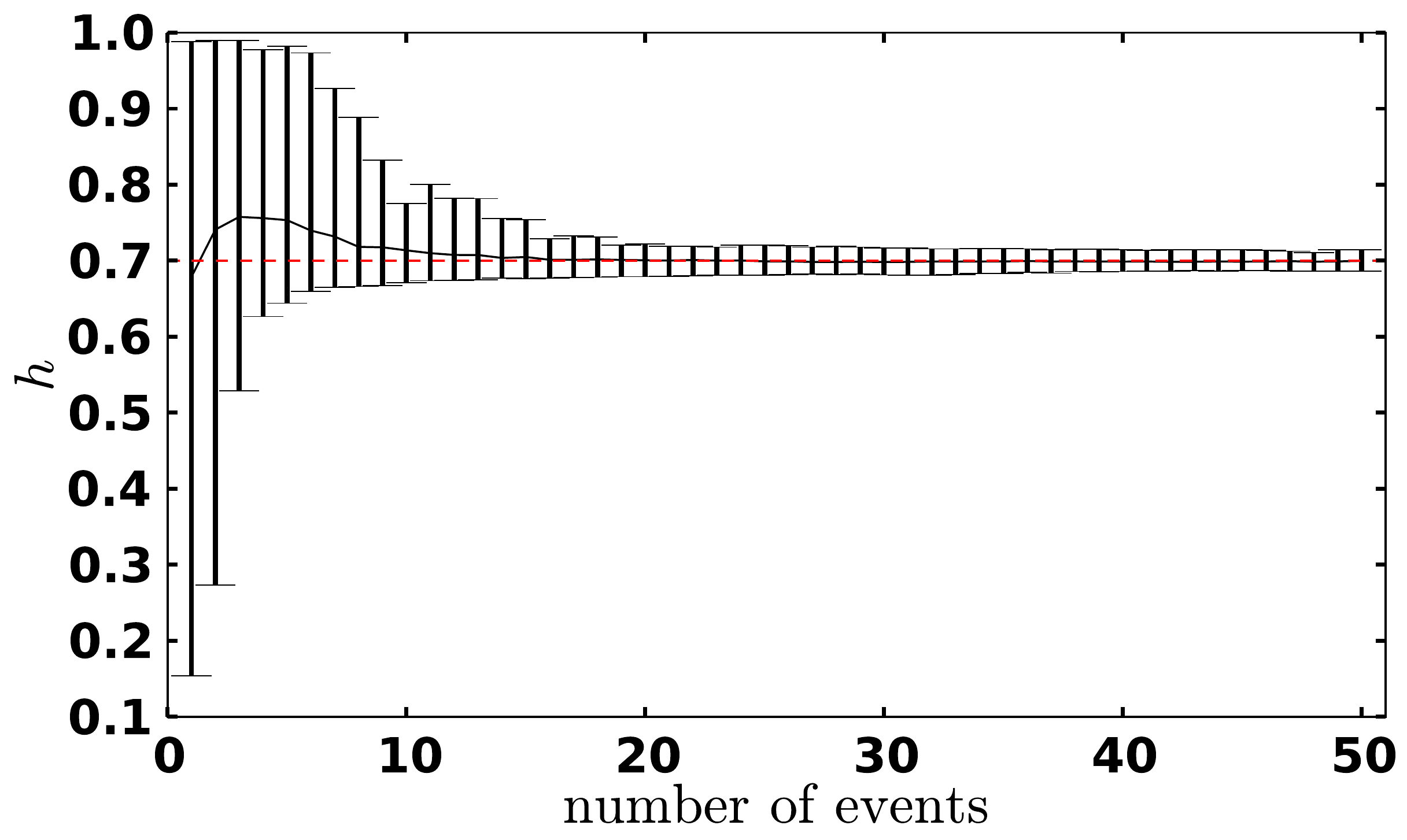}
    \caption{Forecasted determination of the Hubble constant $H_0$ with dark siren events with redshift inferred from galaxy
    catalogs. From \citet{DelPozzo:2011yh}.}
    \label{fig:H0black}
  \end{center}
\end{figure}

Note that as suggested in the original paper \citep{Schutz:1986gp}, a determination of $H_0$ is also possible \emph{without} an electromagnetic counterpart by correlating the distance
measure and sky-localization from GW detectors with galaxy catalogs and associating to the GW events the redshift
of all of the galaxies present in the localized region.
In \cite{DelPozzo:2011yh} it was shown that it will be possible to determinethe Hubble-Lema\^itre constant with
a precision of few \% after 50 \emph{dark sirens} detections, i.e., GW events
without the concurrent presence of electromagnetic transient,
see Figure~\ref{fig:H0black}. In a region of 10 degrees squared, say, $\sim 10^4$ galaxies are expected to be present within a distance up to
$\sim 500$ Mpc, and even if
galaxy catalogs can encompass most of the stellar mass present in the localized
region, and photometric redshift determinations are available (see
\cite{Soares-Santos:2019irc} for an implementation of the idea with a recent binary black hole detection), the number of candidate galaxies will induce a large
error in the final measurement which be counteracted only by combining large numbers of dark sirens.

\subsection{Future detectors}

Beyond the existent LIGOs and Virgo observatories, which are in their \emph{advanced} phase,
 there are plans to build \emph{third generation} detectors,
with the advantage to be able to push their frequency reach down to the $Hz$,
allowing to accumulate much more signals, since the GW amplitude in the
frequency domain scales according to
$\tilde h(f)\propto v^2(f){\dot f}^{-1/2}\sim f^{-7/6}$.

With the third generation detectors Einstein Telescope (ET) \citep{Punturo:2010zz}
and Cosmic Explorer (CE) \citep{Evans:2016mbw} sources at $z\sim 2$ for binary neutron
star signals, and even larger for binary black holes will be accessible,
enabling to accumulate
much more statistics to improve the precision on post-Newtonian and cosmological parameters,
with $O(1000)$ events per month expected.

ET is planned to consist of a three 10-km long Michelson
interferometers arranged in an equilateral triangle to be built underground to
minimize seismic and Newtonian noise.
CE has a similar design, but a L-shape with longer (40 km) arms, offering, like
ET an order of magnitude increase in sensitivity and a wider band extending down
to a few Hertz.

On the astrophysics side it is worth noticing that the number of detectable
sources increases with the observable volume and at low redshift an increase by
a factor $x$ in distance reach implies an $x^3$ enhancement of the number
of sources, but in cosmology
the volume stops increasing with the cube of the distance for large distances,
which has important consequences for the rate of detections.

On general grounds the rate of detected mergers $R_m$ per redshift
can be expressed in terms of the comoving density of mergers
\be
R_m(z_m)\equiv \frac{d N_m}{dt_o dz_m}=\frac{dN_m}{dV_cdt_m}\frac{dV_c}{dz}
\frac 1{1+z_m}\equiv \frac 1{1+z_m}\frac{dV_c}{dz} {\mathcal R}_m(z_m)\,,
\ee
where in the last passage we have defined the comoving volume density rate ${\mathcal R}_m$
of mergers and in the previous one we have used that $dt_o/dt_m=(1+z_m)$.
The comoving density of mergers ${\mathcal R}_m$ is not constant in time and
its modelization
is an active and difficult field of research. However, the main dependence
on red-shift of $R_m$ is actually given by the volume differential factor
$dV_c/dz=4\pi D_c^2dD_c/dz$, with $D_c(z)=\int_0^z H^{-1}(z')dz'$.

\begin{figure}
  \begin{center}
    \includegraphics[width=.65\linewidth]{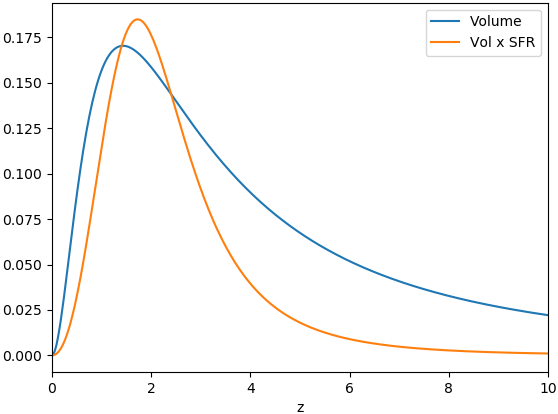}
    \caption{Bare volume factor with respect to redshift (blue) and volume factor times the
realistic star formation rate of \citet{Madau:2014bja}.}
\label{fig:ratevol}
\end{center}

\end{figure}

In Figure~\ref{fig:ratevol} we take the rate of star formation ${\mathcal R}_{sfr}$ from
\cite{Madau:2014bja}:
\be
{\mathcal R}_{sfr}(z)=K \frac{(1+z)^\alpha}{1+\pa{\frac{(1+x)}C}^\beta}
\ee
(with $\alpha=2.7$, $\beta=5.6$, $C=2.9$) and by making the very crude approximation of
equating it to the compact object density merge rate at the same redshift one can show how it affects the detectable merger rates, see Figure~\ref{fig:ratevol}.

Despite some qualitative change by the inclusion of the star formation rate, one
can see that the volume density peaks at around $z\sim 2$ and we expect the detectable merge rate also peak around $z\sim 2$. By collecting $O(10^4)$ events it will
be indeed possible to measure the star formation/merger rate \citep{Vitale:2018yhm} and discriminate among late time cosmic acceleration models \citep{Mendonca:2019yfo}.

\begin{figure}
  \begin{center}
    \includegraphics[width=.7\linewidth]{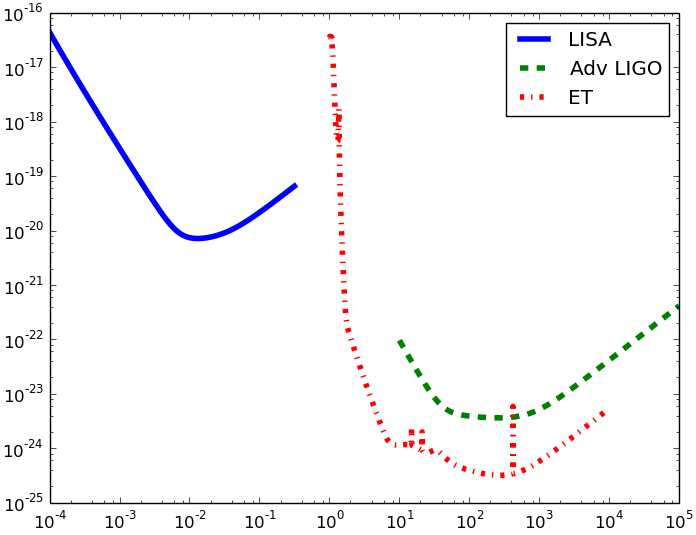}
    \caption{Planned noise curves for eLISA, Einstein Telescope/Cosmic Explorer
and Advanced LIGO. The noise curves have been taken from \cite{Klein:2015hvg} for eLISA, from \cite{ligodcc_evans} for Einstein Telescope and \cite{Harry:2010zz} for
Advanced LIGO.}
    \label{fig:noise}
  \end{center}
\end{figure}

Another GW detector planned for the future is the space interferometer LISA,
which is expected to widen the detection up to $z\sim15$ \citep{Klein:2015hvg,elisaweb}.
The space detector LISA, planned to observe GWs starting from the decade of 2030,
will not be limited in the low frequency region by terrestrial noise and will have a sensitive frequency
band in the region $10^{-3} - 10^{-1}$ Hz, complementing earth-based detectors.
Signals will be much longer: from equation~(\ref{eq:fdot}) it results
that the time $\Delta t(f)$ for the GW signal to evolve from an instantaneous
frequency $f$ to coalescence is given by
\be
\Delta t(f)=\frac{5M_c}{256}\pa{\pi G_NM_cf}^{-8/3} \,,
\ee
thus showing that LISA will have many overlapping sources of GWs.
Another consequence of the opening a low frequency window (a factor $10^4$ lower than LIGO) is the possibility to
observe systems up to a mass of $\sim 10^6M_\odot$ (i.e.~$10^4$ higher than LIGO)
hence starting to access the realm of supermassive black holes.
See Figure~\ref{fig:noise} for the planned noise curves for eLISA, Einstein Telescope/Cosmic Explorer
and Advanced LIGO.

%%%%%%%%%%%%%%%%%%%%%%%%%%%%%%%%%%%%%%%%%%%%%%%%%%%%%%%%%%%%%%%%
\section*{Acknowledgments}

It is a pleasure to thank Raul Abramo and David Camarena for useful feedback.
VM thanks CNPq and FAPES for partial financial support.
RR thanks CNPq and FAPESP for partial financial support.
RS thanks CNPq for partial financial support.
RS used data obtained from the Gravitational Wave Open Science Center (https://www.gw-openscience.org), a service of LIGO Laboratory, the LIGO Scientific Collaboration and the Virgo Collaboration. LIGO is funded by the U.S. National Science Foundation. Virgo is funded by the French Centre National de Recherche Scientifique (CNRS), the Italian Istituto Nazionale della Fisica Nucleare (INFN) and the Dutch Nikhef, with contributions by Polish and Hungarian institutes.

\bibliographystyle{plainnat}
\bibliography{references,biblio-RP,ref}

\end{document}